\documentclass{article}

\usepackage{arxiv}

\usepackage[utf8]{inputenc} 
\usepackage[T1]{fontenc}    
\usepackage{url}            
\usepackage{booktabs}       
\usepackage{amsfonts}       
\usepackage{nicefrac}       
\usepackage{microtype}      
\usepackage{lipsum}		
\usepackage{graphicx}
\usepackage{lineno}
\usepackage[utf8]{inputenc}
\usepackage{lipsum}
\usepackage{multicol}
\usepackage[table]{xcolor}
\usepackage{tikz}
\usepackage{amsmath}
\usepackage{amsthm}
\usepackage{flushend}
\RequirePackage{doi}
\usepackage{lineno,hyperref}
\usepackage[utf8]{inputenc}
\usepackage{lipsum}
\usepackage{multicol}
\usepackage[table]{xcolor}
\usepackage{tikz}
\usepackage{amsmath}
\usepackage{amsthm}
\usepackage{flushend}
\usepackage{multirow}
\usepackage{algorithm}
\usepackage{algpseudocode}
\usepackage{amsmath}
\usepackage{longtable}

\newcommand{\RN}[1]{%
	\textup{\uppercase\expandafter{\romannumeral#1}}%
}

\definecolor{bar}{rgb}{0.557,0.663,0.859}
\definecolor{back}{rgb}{0.851,0.882,0.949}

\title{Towards Benchmark Datasets for Machine Learning Based Website Phishing Detection: An experimental study}

\author{ {\hspace{1mm}Abdelhakim Hannousse}
	\\
	Department of Computer Science\\
	Universté 8 Mai 1945, Guelma\\
	 BP 401, Guelma 24000, Algeria\\
	\texttt{hannousse.abdelhakim@univ-guelma.dz} \\
	\And
	{\hspace{1mm}Salima Yahiouche} \\
	Department of Computer Science\\
	LRS laboratory, Badji Mokhtar University\\
	BP 12, Annaba 23000, Algeria\\
	\texttt{yahiouche.salima@univ-annaba.dz} \\
}

\renewcommand{\headeright}{}

\begin{document}
\maketitle

\begin{abstract}
The increasing popularity of the Internet led to a substantial growth of e-commerce. However, such activities have main security challenges primary caused by cyberfraud
and identity theft. Therefore, checking the legitimacy of visited web pages is a crucial task to secure costumers' identities and prevent phishing attacks. The use of machine learning is widely recognized as a promising solution. The literature is rich with studies that use machine learning techniques for website phishing detection. However, their findings are dataset dependent and are far away from generalization. Two main reasons for this unfortunate state are the impracticable replication and absence of appropriate benchmark datasets for fair evaluation of systems. Moreover, phishing tactics are continuously evolving and proposed systems are not following those rapid changes. 
In this paper, we present a general scheme for building reproducible and extensible datasets for website phishing detection. The aim is to (1) enable comparison of systems using different features, (2) overtake the short-lived nature of phishing websites, and (3) keep track of the evolution of phishing tactics. For experimenting the proposed scheme, we start by adopting a refined classification of website phishing features and we systematically select a total of 87 commonly recognized ones, we classify them, and we made them subjects for relevance and runtime analysis. We use the collected set of features to build a dataset in light of the proposed scheme. Thereafter, we use a conceptual replication approach to check the genericity of former findings for the built dataset. Specifically, we evaluate the performance of classifiers on individual classes and on combinations of classes, we investigate different combinations of models, and we explore the effects of filter and wrapper methods on the selection of discriminative features. The results show that Random Forest is the most predictive classifier. Features gathered from external services are found the most discriminative where features extracted from web page contents are found less distinguishing. Besides external service based features, some web page content features are found time consuming and not suitable for runtime detection. The use of hybrid features provided the best accuracy score of 96.61\%. By investigating different feature selection methods, filter-based ranking together with incremental removal of less important features improved the performance up to 96.83\% better than wrapper methods.
\end{abstract}

\keywords{Website phishing attacks\and Machine learning\and Dataset benchmarking\and Information security}

\section{Introduction}
Phishing has been recognized as the easiest and widespread cybercriminality threat. Hackers do not need to crack any complex cypher code neither breach a hard firewall. Instead, they simply send emotional, critical or sensible e-mails urging recipients to introduce their personal credentials by clicking on a link. Recipients are then redirected to fake web pages that look very similar to those targeted authentic websites. Consequently, recipients are trapped in fake websites like fishes.
Recently, hackers start doing their jobs very professionally, the recent phishing activity trends report~\cite{APWG2020} showed that 78\% of all phishing websites use SSL protection that was exclusively used by authentic websites. Wandera stated in its 2020 Mobile Threat Landscape Report~\cite{Wandera2020} that a new phishing website launches every 20 seconds. All these facts advocate deep research studies on the detection and prevention of such cybercriminality attacks. 

The detection of phishing attacks is mainly mapped into a classification problem. Therefore, machine learning techniques are considered as  promising solutions. However, three major aspects need to be considered when adopting such techniques: (1) selection of efficient classifiers, (2) usage of distinguishing features, and (3) collection of representative dataset samples for training. 
Machine learning based systems developed for phishing detection are mostly classified into two main categories: content-based and URL-based systems. In the former case, phishing is detected by active or passive examination of the content of visited web pages. In the latter case, only URLs of visited web pages are examined. 

%
Recently, Das et al. have conducted a systematic review of published studies in the period 2010-2017 that used machine learning techniques for website phishing detection~\cite{Das2020}. It was found that SVM is the most used classifier for content-based phishing detection, followed by Logistic Regression, Decision Tree and Naïve Bayes. For URL-based detection, Decision Tree came first, followed by Random Forest, SVM, and Logistic Regression. This shows divergent views on which classifier is more predictive for each class of features. However, the results in~\cite{Das2020} are only based on statistical analysis and do not amount to be a decisive fact. 


Moreover, a large number of distinct features are proposed for feeding classifiers. Das et al. rated features according to the number of studies stating them~\cite{Das2020}. The use of IP addresses instead of domain names was found the most adopted feature for URL-based approaches, followed by length of URLs, frequency of special characters and number of dots. Most used features extracted from web page contents were the number of internal/external links, term frequency, HTML tag attributes and the number of various tag types.
Chiew et al. have advocated the exclusive use of URLs for phishing detection to reduce the risk of accessing  harmful web pages~\cite{Chiew2018}. Jain et al. used only content-based features and suggested the exclusion of external-based features for their time consuming~\cite{Jain:2019}. This shows again the lack of common agreement on which class of features is suitable for the detection of phishing websites.  
Dou et al. in~\cite{Dou2017} stated that despite being discriminative, selected features must also be robust. Therefore, even if features are distinguishing enough, they must not be manipulated easily by hackers. In addition, the time taken for their extraction is also very important to avoid any delay in the real time detection of instances. 

Furthermore, the majority of contemporary studies use self-collected datasets from different sources. The UCI repository\footnote{UCI repository: \url{https://archive.ics.uci.edu/}} also provides a pre-elaborated dataset for phishing detection. However, those datasets are not suitable for replication and experimentation with new features. This is due to two primary causes: 

\begin{enumerate}
	\item Absence of URLs used for building the datasets. 
	\item The short-lived nature of phishing websites. 
\end{enumerate}

The above causes prevent the quality analysis of used datasets and make the comparison of systems within a same dataset quasi impossible. Hence, proposed systems can easily be considered as dataset dependent and the findings cannot be generalized. Therefore, the presence of benchmark datasets becomes a necessity. 
To alleviate this problem, the contribution of the present paper are as follows:

\begin{enumerate}
	\item Propose and experiment a construction scheme of reproducible and extensible datasets for web page phishing detection. A dataset is built in light of the proposed scheme which may serve as a benchmark for comparing systems. 
	\item Propose a refined classification of features for phishing detection. Commonly used features are  systematically collected from recent reviews and classified following the proposed classification scheme.
	\item Examine former findings regarding the performance of classifiers and the suitability of features for runtime detection. 
	\item Experiment different feature selection methods.
	\item Identify the model that provides the best accuracy for the collected dataset.
\end{enumerate}

The remaining of the paper is organized as follows. Section~\ref{sec:rw} discusses related works. Section~\ref{sec:method} describes the proposed scheme for the collection of reproducible and extensible datasets for website phishing detection. Section~\ref{sec:data} experiments the proposed guidelines through the collection of a dataset sample that is used for the experimentations conducted in the study. Section~\ref{sec:process} describes the methodology adopted for conducting the experiments within the collected dataset. Section~\ref{sec:results} presents the obtained results, section~\ref{sec:discussion} discusses the results and draws some conclusions, and section~\ref{sec:conclusion} concludes the paper and shows some perspectives. 

\section{Related Work}
\label{sec:rw}
\begin{figure}[h]
	\centering
	\includegraphics[width=.6\textwidth]{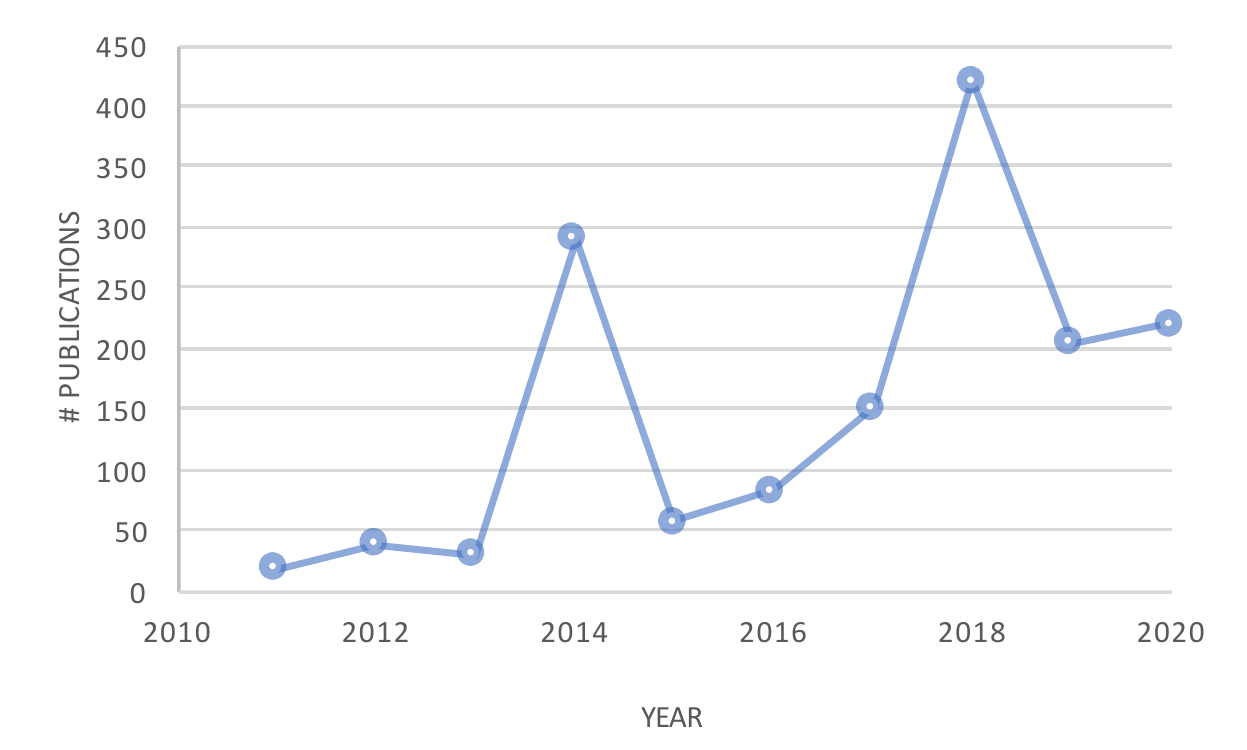}
	\caption{Number of publications per year focusing on the use of machine learning for the detection of phishing websites. The results are taken the 16th of September 2020 from \url{ https://app.dimensions.ai} using the following query search: "\textit{website}" \textbf{AND} "\textit{phishing detection}" \textbf{AND} "\textit{machine learning}".}     
	\label{fig:publications} 
\end{figure}

The literature is rich with studies proposing machine learning techniques for website phishing detection, and the number of papers increases continuously every year. Figure~\ref{fig:publications} shows the number of publications per year in the last decade. The results has been retrieved from Dimensions research grants database\footnote{Dimensions research grants database: \url{ https://app.dimensions.ai}}. 

Dou et al. in~\cite{Dou2017} have conducted a systematic review on machine learning based phishing detection systems. The authors noticed that runtime performance was neglected by most systems, URL and content-based features are the most commonly used, and studies that incorporate more features have better performance results. Das et al. in~\cite{Das2020} have conducted a similar review but also ranked classifiers regarding the number of studies adopting them. However, those conclusions are only derived from statistical analysis of examined studies. Contrary to these works, the aim of the present study is to check the validity of those results experimentally. 

Few are the studies that use merely content-based features. Jain et al. in~\cite{Jain:2019} extracted 12 features based on the nature of hyperlinks in the content of web pages. The experimentation of several classifiers showed that logistic regression has the best accuracy score. However, most of found studies in the literature use hybrid set of features. Jain et al. in ~\cite{Jain2018a} used an SVM trained on a mixed set of URL and third party based features. The experimentation showed more than 90\% of accuracy in detecting phishing websites.
Srinivasa-Rao et al.~\cite{SrinivasaRao2019} used random forest classifier trained on 35 URL based features and term frequency and inverse document frequency (TF-IDF) based features extracted from the content of web pages. The model has achieved an accuracy up to 98.25\%.
Shirazi et al.~\cite{Shirazi2018} used only 7 features extracted from both URLs and web page contents. The authors experimented the performance of several classifiers. Best results were obtained by a gradient boosting classifier with 97\% of accuracy.
Zaini et al.~\cite{Zaini2019} selected 15 features from different classes. They experimented five machine learning classifiers and found that random forest achieved the highest detection accuracy of 94.79\%.
In this study, we experiment the importance of individual class of features and all the different possible combinations of feature classes.

Several techniques are used for the selection of discriminative features. These techniques include the use of ranking filters and wrapper algorithms. 
Korkmaz et al.~\cite{Korkmaz2020} identified 120 features used in existing studies. Those features are classified into three separate classes. Like previously discussed works~\cite{Dou2017, Das2020}, features were compared regarding the number of studies adopting them. Additionally, features were ranked using chi-square, f-statistic, mutual information, and Pearson correlation but only one class of features were tested. 
Rajab et al.~\cite{Rajab2017} proposed a new ranking metric of features by combining estimated ranks by information gain and chi-square algorithms. The authors experimented the results within the UCI dataset and two classifiers, namely C4.5 and JRIP. The results showed an improvement in the accuracy of the C4.5 classifier that attains more than 95.5\%.
Nagaraj et al.~\cite{Nagaraj2018} started with 21 features and reduced the number to 6 using a threshold adopted for the outputs of Boruta wrapper algorithm ranking~\cite{Kursa2010}.
In this paper, we also experiment the selection of features using ranking filters like in~\cite{Korkmaz2020} but we take into account all classes of features. We also experiment different wrapper algorithms for features selection including the Boruta wrapper algorithm used in~\cite{Nagaraj2018}.


Some studies tried to combine classifiers to improve the accuracy in the detection of phishing websites. Nagaraj et al.~\cite{Nagaraj2018} developed a two fold ensemble learner by using the outputs of a random forest classifier to feed a neural network classifier. 
Sameen et al.~\cite{Sameen2020} designed an ensemble machine learning based on majority voting. The authors of~\cite{Sameen2020} adopted 17 features extracted from URLs and used 10 machine learning classifiers within a multi-threaded technique to speed up the process and enable real time detection. The proposed model achieved 98\% in term of accuracy.
In this study, we experiment the effect of different combinations of classifiers including stacking as in~\cite{Nagaraj2018} and majority voting as in~\cite{Sameen2020}. 

\section{Scheme for constructing reproducible and extensible datasets}
\label{sec:method}
Benchmark datasets for website phishing detection are not available. This is due to the fact that phishing websites are short-time living and dead URLs cannot be used in content-based analysis. Besides, most available datasets contain only the values of experimented features without referring to used URLs. This prevents replication and experimenting those datasets with different features. Moreover, El-Aassal et al.~\cite{ElAassal2020} evaluated the impact of the ratio between phishing and legitimate samples in datasets. The experiments showed that imbalanced datasets may decline the performance of classifiers from 5.9\% to 42\% in term of F-1 score.
For such reasons, we propose the following guidelines for the construction of reproducible and extensible datasets for website phishing detection. Those datasets may serve as benchmarks in the domain. The following proposed guidelines are deduced from current common and best practices in the field:

\begin{description}
	\item [\textbf{G1}:] Start by collecting URLs preferably from different sources. It is found in~\cite{Dou2017} that a large number of studies collect legitimate websites from Alexa\footnote{Alexa website: \url{https://www.alexa.com/}}. However, Alexa only proposes top ranked domains without referring neither to sub-domains nor paths. Therefore, for the homogeneity of the dataset, one cannot use Alexa lists directly especially when features concerned with subdomains and paths are adopted. To deal with this issue, it is proposed in~\cite{Das2020} to use the list of top domains provided by Alexa as seeds to crawl more generative URLs. For the diversity of URLs, one may use URLs from Alexa categories as used in~\cite{Das2020, ElAassal2020}. 
	
	\item[\textbf{G2}:] Collected URLs need to be preprocessed. The preprocessing step includes: (1) remove duplicate URLs, (2) avoid using much URLs from the same domain since those URLs are likely to have similar feature values. This enables obtaining more representative samples.
	
	\item[\textbf{G3}:] When extracting features, keep track of used URLs either by using them as an index for the dataset or saving them in a separate file and made them available for users. This enables the reproduction of URL-based features and extend the main dataset with more features.
	
	\item[\textbf{G4}:] To overcome the short-lived nature of phishing web pages, one can use available tools to generate the Document Object Model (DOM) tree of pages and save them in the dataset or in a separate dataset indexed by URLs. This enables experimenting and extending the main dataset with more content-based features even with dead links. 
	
	\item[\textbf{G5}:] Final datasets should be shuffled and balanced evading the issue reported in~\cite{ElAassal2020} regarding the impact of unbalanced datasets on the performance of classifiers.  
	
	\item[\textbf{G6}:] Collected datasets should be dated. Therefore,  systems can be tested on datasets collected in different and reasonable time spans checking the efficiency of models and the significance of adopted features. This also enables keeping track of phishing tactics evolution over time.
	
\end{description} 

\section{Dataset preparation}
\label{sec:data}
\begin{figure}[h]
	\centering
	\includegraphics[width=1\textwidth]{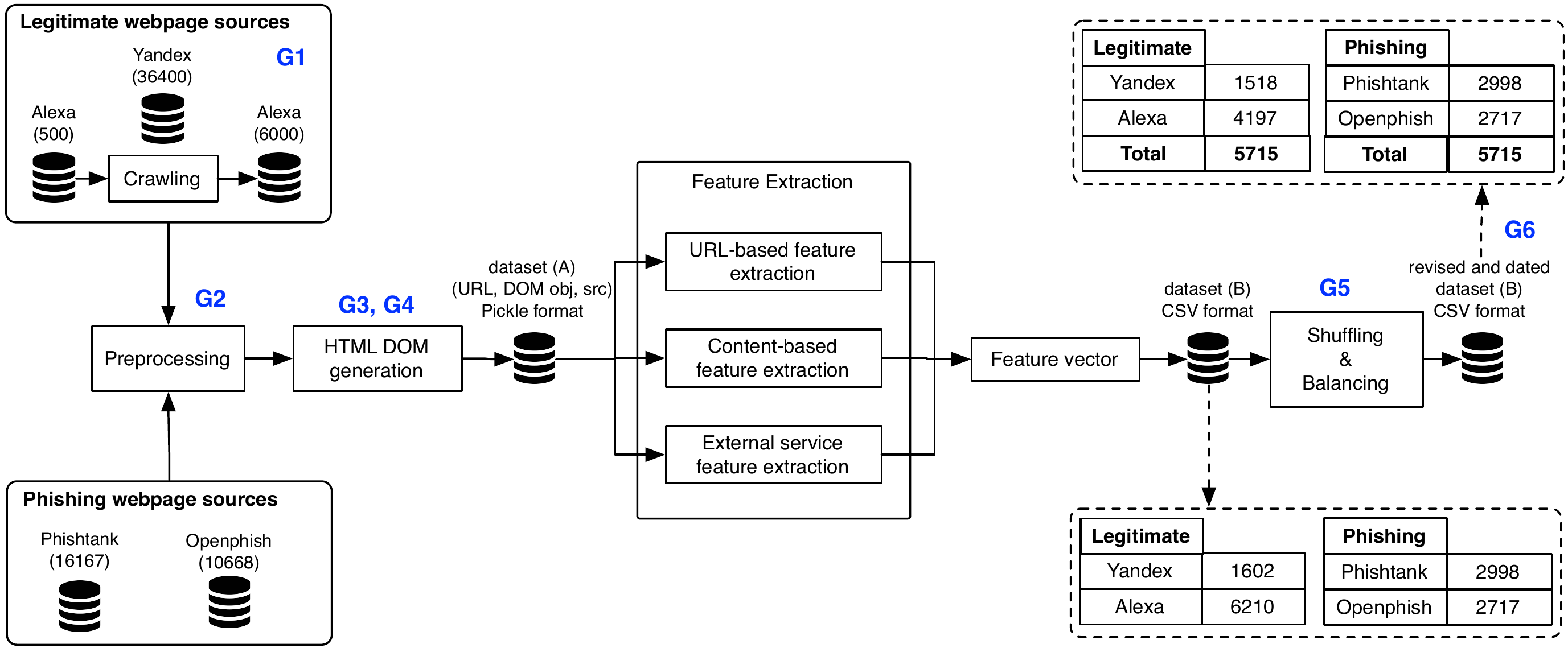}
	\caption{Dataset construction process.}     
	\label{fig:dataset} 
\end{figure}

In this section, we use the proposed scheme in section~\ref{sec:method} for the collection of a dataset that is used later for the experiments conducted in this study. Hence, we collected a reasonable in size dataset of 11430 phishing and legitimate URLs. Figure~\ref{fig:dataset} depicts the overall process adopted for the construction of the dataset.

\subsection{URL collection}
Legitimate web page URLs are collected from Alexa and Yandex~\footnote{Yandex search engine: \url{https://tech.yandex.com.tr/xml/}}. In the first case, the first 500 top domains provided by Alexa for free are used. Those top domains are adopted as seeds for crawling 12 URLs per domain. Therefore, we ended up with 6000 URLs from Alexa. In the second case, we used still active legitimate URLs provided by Sahingoz et al.~\cite{Sahingoz2019}. Those legitimate URLs were originally collected from Yandex. For phishing URLs, both lists of URLs provided by Phishtank\footnote{Phishtank website: \url{https://www.phishtank.com/}} and Openphish\footnote{Openphish website: \url{https://openphish.com/}} are used. Collected URLs are preprocessed where all duplicate and dead URLs are removed and maximum of 12 URLs with the same domain name are kept. Approved URLs are stored keeping track of their sources (i.e., Alexa, Phishtank, etc.). Afterwards, HTML Document Object Model (DOM) trees are generated for all approved URLs and stored in a supplementary dataset (\textit{dataset A} in Figure~\ref{fig:dataset}).  
HTML DOM Parser for Python\footnote{HTML DOM Parser package for Python: \url{http://thehtmldom.sourceforge.net/}} is used for this task and \textit{dataset A} is stored in a separate pickle file. DOM trees enable the exploration of the structures and contents of URL web pages in an efficient way, where pickle\footnote{Python pickle module: \url{https://docs.python.org/3/library/pickle.html}} is a module that implements binary protocols for serializing and de-serializing complex Python objects. 


\subsection{Feature extraction}
As the aim of the study is to experiment a maximum number of popular features, we examined recent reviews analyzing the usage frequency of features used for website phishing detection. The list of examined reviews are given in Table~\ref{tab:reviews}. The examination showed that several features are given with different names and/or different measures (e.g., ratio or absolute number). To alleviate these issues, we iteratively executed the following two-step process and we ended up with 87 features: (1) identify and remove duplicate features, and (2) adopt a single and commonly used measurement for each feature.

\begin{small}
	\begin{table}[h]
		\centering
		\caption{List of reviews used for collecting features}
		\label{tab:reviews}       
		\begin{tabular}{lcc}
			\hline\noalign{\smallskip}
			Reviews & Year & \#Features \\
			\noalign{\smallskip}\hline\noalign{\smallskip}
			Korkmaz et al. \cite{Korkmaz2020} &2020&	120	\\
			Das et al.~\cite{Das2020} & 2020 & 88\\
			El-Assal et al.~\cite{ElAassal2020}& 2020 & 83\\
			Althobaiti et al.~\cite{Althobaiti2019} & 2019 &  40\\
			\noalign{\smallskip}\hline
		\end{tabular}
	\end{table}
\end{small}

\begin{figure}[h]
	\centering
	\includegraphics[width=.9\textwidth]{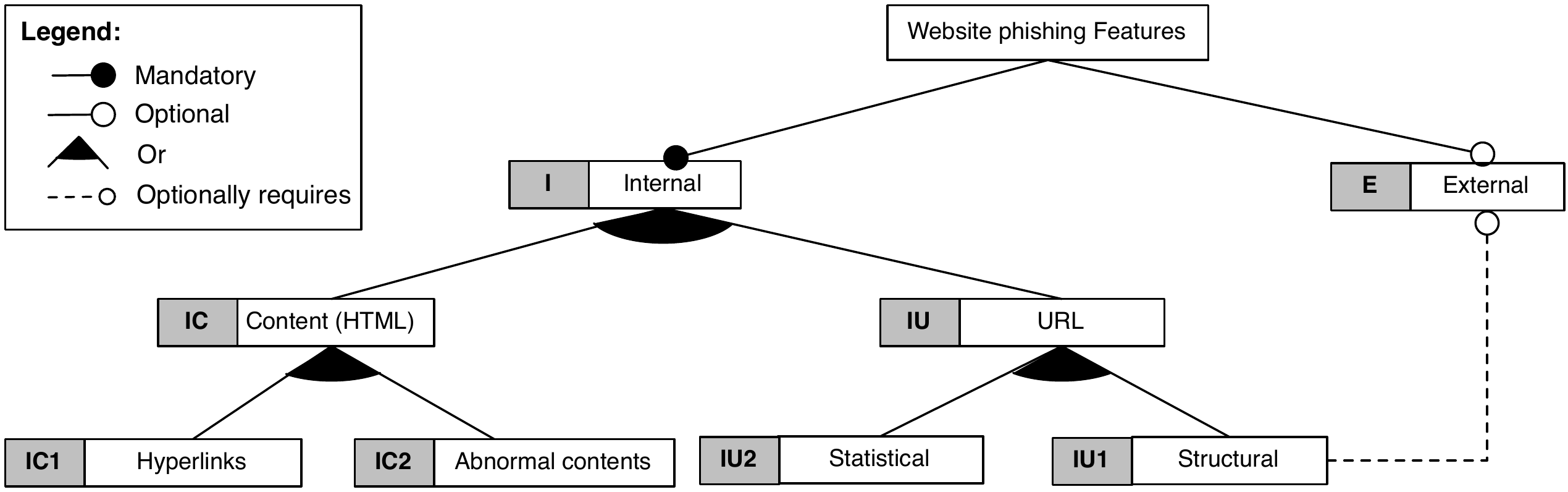}
	\caption{Classification of antiphishing features.}     
	\label{fig:features} 
\end{figure}

%
The examination of the literature showed different classification schemes of website phishing features. Figure~\ref{fig:features} depicts a feature diagram showing common, alternative and optional class of features adopted by existing antiphishing systems. This classification is used to categorize the features examined in this study.

Contemporary machine learning-based antiphishing systems are found using internal with or without external  features. This is modeled in Figure~\ref{fig:features} by a mandatory relationship with internal features and an optional relationship with external features. Internal features (\textbf{I}) are extracted either directly from URLs (\textbf{IU}) or from HTML contents of their corespondent web pages (\textbf{IC}). Hybrid systems use collection of features from both \textbf{IC} and \textbf{IU}. This is modeled by the \textit{Or} relationship in Figure~\ref{fig:features}. 
External features (\textbf{E}) are used as complementary and often considered as URL-based features. This is due to the fact that they are attained through querying external services with URLs or URL domains. 

In this study, the feature extraction process is fully automated by developing a Python script that takes a web page URL as input and generates its correspondent feature vector of size 87x1. For keeping track of the source URL and made it available for further usage, we used the input URL as an index for the generated dataset.
The subsequent subsections describe the different features of each class presented in Figure~\ref{fig:features}.

\subsection*{URL-based features}
URL-based features (\textbf{IU}) are obtained by simply analyzing the text of URLs. Features of this class can also be divided into \textit{structural} and \textit{statistical} features. Structural-based features (\textbf{IU1}) are concerned with the presence, position and nature of URL base elements (i.e., protocol, domain, subdomans, path, port, and top level domain). Examples of such features are the presence of ports, use of '\texttt{https}' protocol, and position of the top level domain (TLD). 
Some structural-based features may require the use of external services for their estimation. Examples of such features are checking for the suspiciousness of TLDs and brand domains. Alternatively, predefined lists of required data can be collected and used for each feature. This justifies the \textit{optionally requires} relationship between structural and external-based features in Figure~\ref{fig:features}. 
Statistical-based features (\textbf{IU2}) are concerned with the number or distribution of URL base elements, specific words, or characters in the text of URLs. Examples of those features are the number of dots and subdomains and length of words.
The set of experimented features of this class are given in Table~\ref{tab:urlfe} together with their type of values, subclasses, and intuitions behind their adoption. In the table, IU1$^{*}$ refers to structural features that may require external services for their estimation. 

\begin{small}
	\begin{longtable}{l|p{.7in}|l|c|p{3.9in}}
		\caption{URL-based features for website phishing detection}
		\label{tab:urlfe}   
		\endfirsthead
		\endhead
		\hline\noalign{\smallskip}
		Index & Feature & Type & Class & Intuition\\
		\noalign{\smallskip}\hline\noalign{\smallskip}
		\textbf{f1-2}& URL parts lengths & Int & IU2 & Long URLs are used to hide real domains and subdomains. We consider: full URL length (\textbf{f1}) and hostname length (\textbf{f2}).\\
		\noalign{\smallskip}\hline\noalign{\smallskip}
		\textbf{f3} & IP & 0/1 & IU1 & IP addresses are used in hostnames to hide the identity of websites. IPs can also be used without dots or hexadecimal encoded. Presence of IPs in any format in hostnames is considered as phishing indicator.\\
		\noalign{\smallskip}\hline\noalign{\smallskip}
		\textbf{f4-20} & Special Characters & Int & IU2 & Special characters are used to deceive novice users of real domains and subdomains. We consider the number of occurrences of the following characters:  '\texttt{.}' (\textbf{f4}), '\texttt{-}' (\textbf{f5}), '\texttt{@}' (\textbf{f6})  , '\texttt{?}' (\textbf{f7}), '\texttt{\&}' (\textbf{f8}), '\texttt{$\mid$}' (\textbf{f9}), '\texttt{=}' (\textbf{f10}), '\texttt{\_}' (\textbf{f11}), '\texttt{\~{}}' (\textbf{f12}), '\texttt{\%}' (\textbf{f13}), '\texttt{/}' (\textbf{f14}), '\texttt{*}' (\textbf{f15}), '\texttt{:}' (\textbf{f16}), '\texttt{,}' (\textbf{f17}), '\texttt{;}' (\textbf{f18}), '\texttt{\$}' (\textbf{f19}), '\texttt{\%20}' or space (\textbf{f20}).\\
		\noalign{\smallskip}\hline\noalign{\smallskip}
		\textbf{f21-24} & \#common terms & Int & IU2 & Common terms in URLs such as '\texttt{www}' (\textbf{f21}), '\texttt{.com}' (\textbf{f22}),  '\texttt{http}' (\textbf{f23}) and '\texttt{//}' (\textbf{f24}) are used only once in legitimate URLs where it is observed that they are used more than once in phishing URLs.\\
		\noalign{\smallskip}\hline\noalign{\smallskip}
		\textbf{f25} & HTTPS ~~~~~~~~ token& 0/1 & IU1& Most phishing websites do not provide any security facilities compared with legitimate ones. Thus, the use of HTTPS is a legitimacy indicator.\\
		\noalign{\smallskip}\hline\noalign{\smallskip}
		\textbf{f26-27} & Ratio of ~~~~~~~~ digits & Float & IU2 & High number of digits in URLs is considered as a phishing indicator. We consider ratio of digits in full URLs (\textbf{f26}) and hostnames (\textbf{f27}).\\
		\noalign{\smallskip}\hline\noalign{\smallskip}	
		\textbf{f28}& Punycode& 0/1 & IU1 & Punycode is used in domain names to replace some ASCIIs with Unicode characters. URLs will then look legitimate where they refer to different websites. URLs with punycodes are considered phishing.\\
		\noalign{\smallskip}\hline\noalign{\smallskip}
		\textbf{f29} & Port& 0/1 & IU1 & Port numbers are rarely used in legitimate URLs. Therefore, URLs with port indicator are considered phishing.\\
		\noalign{\smallskip}\hline\noalign{\smallskip}
		\textbf{f30-31} & TLD ~~~~~~~ position & 0/1 & IU1 & In well-formed URLs, top-level domains (TLDs) appear only before the path. When TLDs appear in the path (\textbf{f30}) or in the subdomain part (\textbf{f31}), the URL is considered phishing.\\
		\noalign{\smallskip}\hline\noalign{\smallskip}
		\textbf{f32} & Abnormal subdomains & 0/1 & IU1 & Phishing URLs may use the following pattern '\texttt{w[w]?[0-9]*}' instead of '\texttt{www}' to deceive users. Thus, URLs with subdomains matching such pattern are considered phishing.\\
		\noalign{\smallskip}\hline\noalign{\smallskip}
		\textbf{f33} & \#subdomains & Int & IU2 & Phishing URLs use more number of subdomains compared with legitimate ones. Thus, the number of subdomains is a phishing feature.\\
		\noalign{\smallskip}\hline\noalign{\smallskip}\
		\textbf{f34}& Prefix Suffix& 0/1 & IU1 & Prefixes and suffixes separated by "-" are used in domain names to make users feel that they are dealing with legitimate pages. Thus, when "-"  is found in domain names, the URL is considered phishing. \\
		\noalign{\smallskip}\hline\noalign{\smallskip}
		\textbf{f35} & Random ~~~ domains& 0/1 &IU1 & Phishing URLs use words formed from random characters. Therefore, domain names are checked for randomness. \\
		\noalign{\smallskip}\hline\noalign{\smallskip}
		\textbf{f36}& Shortening service& 0/1 & IU1$^{*}$ & URL shortening service is used to indicate short URLs that serve as a redirect to other long and complex URLs. This service can be used in phishing to hide the name of real hosts. Therefore, the use of shortening service is considered as a phishing indicator. The list of shortners are extracted from URLTeam tracker available at \url{https://www.archiveteam.org/index.php?title=URLTeam#URL_shorteners}. \\
		\noalign{\smallskip}\hline\noalign{\smallskip}
		\textbf{f37} & Path ~~~~~~~~~~ extension& 0/1 & IU1$^{*}$ & Malicious scripts can be added to legitimate pages. Some file extensions used in URL paths may lunch such kind of attacks. Presence of the following malicious path extensions is considered: '\texttt{txt}', '\texttt{exe}', '\texttt{js}'\\
		\noalign{\smallskip}\hline\noalign{\smallskip}
		\textbf{f38-39}& Redirections & Int & IU2 & URL redirection is a technique used to open pages with different URLs than those initially selected by users. This is useful to prevent access to broken links when web pages are moved. URLs can be redirected to pages with the same domain (i.e. internal redirection) or to pages from different domains (i.e., external redirections). However, redirection can also be used for hostile purposes. The number of redirections (\textbf{f38}) and external redirections (\textbf{f39}) are considered as phishing indicators.\\
		\noalign{\smallskip}\hline\noalign{\smallskip}
		\textbf{f40-50}& NLP ~~~~~~~~~~ features& Int & IU2 & Natural language processing and word-raw features are also used in phishing detection. We consider number of words (\textbf{f40}), char repeat (\textbf{f41}), shortest words in URLs (\textbf{f42}), hostnames (\textbf{f43}), and paths (\textbf{f44}), longest words in URLs (\textbf{f45}), hostnames (\textbf{f46}), and paths (\textbf{f47}), average length of words in URLs (\textbf{f48}), hostnames (\textbf{f49}), and paths (\textbf{f50}).\\
		\noalign{\smallskip}\hline\noalign{\smallskip}
		\textbf{f51} & Phish hints & Int & IU2 & Phishing URLs use sensitive words to gain trust on visited web pages. The number of such words in URLs is considered as phishing indicator. \\
		\noalign{\smallskip}\hline\noalign{\smallskip}
		\textbf{f52-54} & Brand ~~~~~~ domains & 0/1 & IU1$^{*}$  & Phishing URLs use brand domain names in different URL parts. The presence of brand names in the domain part is considered as a legitimacy indicator (\textbf{f52}) where their presence in subdomains (\textbf{f53}) or paths (\textbf{f54}) is considered as a phishing indicator. The list of brand domain names are collected from~\url{https://www.101domain.com/}.\\
		\noalign{\smallskip}\hline\noalign{\smallskip}
		\textbf{f55} & Suspicious TLD & 0/1 & IU1$^{*}$  & TLDs are checked for suspiciousness. List of suspicious TLDs, used in this study, are collected from: Spamhaus.org (\url{https://www.spamhaus.org}) and Blue coast system Inc. (\url{https://www.broadcom.com/}).\\
		\noalign{\smallskip}\hline\noalign{\smallskip}
		\textbf{f56} & Statistical report & 0/1 & IU1$^{*}$ & URL domains are checked if their IP addresses match one of top phishing domains. The list is collected from previous works.\\
		\noalign{\smallskip}\hline
	\end{longtable}
\end{small}

\subsection*{Content-based features}
Content-based features (\textbf{IC}) are extracted by loading the web pages of URLs and analyzing their HTML contents. They can be divided into \textit{hyperlink} and \textit{abnormal content} based features. Hyperlink features (\textbf{IC1}) are concerned with the number, status, and nature of hyperlinks (i.e., internal/external) used in HTML tags. Abnormal content features (\textbf{IC2}) are concerned with the identification of suspicious contents or scripts implementing suspicious behaviors. Examples of suspicious contents are the use of empty links and different domain names in the title tag of web pages. Examples of suspicious behaviors are submitting form contents to emails and disabling right clicks. 
Content-based features experimented in this study are given in Table~\ref{tab:ctnfe}.

\begin{small}
	\begin{longtable}{l|p{.7in}|l|c|p{3.9in}}
		\caption{Content-based features for website phishing detection}
		\label{tab:ctnfe}       
		\endfirsthead
		\endhead
		\hline\noalign{\smallskip}
		Index & Feature & Type & Class & Intuition\\
		\noalign{\smallskip}\hline\noalign{\smallskip}
		\textbf{f57} & \#hyperlinks & Int & IC1 &Legitimate websites are supposed to consist of bigger number of pages compared with phishing ones. Therefore, the number of links in URL web page contents is considered for distinguishing phishing websites.\\
		\noalign{\smallskip}\hline\noalign{\smallskip}
		\textbf{f58-59} & Ratio internal/external hyperlinks & Float & IC1 & Legitimate pages usually use hyperlinks with the same base domain of the website while phishing pages use more external hyperlinks pointing to target websites. The ratio of internal (\textbf{f58}) and external (\textbf{f59}) hyperlinks of web pages are considered as phishing indicators.\\
		\noalign{\smallskip}\hline\noalign{\smallskip}
		\textbf{f60} & Ratio null hyperlinks & Float & IC1 & To mimic target pages, the same hyperlinks of legitimate web pages appear in phishing web pages but with empty links. Therefore the ratio of null hyperlinks in tags is used as a phishing indicator.\\
		\noalign{\smallskip}\hline\noalign{\smallskip}
		\textbf{f61} & \#External CSS& Int & IC1 &  Legitimate websites use an internal style or more than one CSS file. Instead, phishing websites use only a sole external CSS file that contains links to CSS files of target websites. Consequently, the number of external CSS files is considered as a phishing indicator.\\
		\noalign{\smallskip}\hline\noalign{\smallskip}
		\textbf{f62-63} & \#Internal/ External redirections & Int & IC1 & Links in phishing web pages may redirect to other legitimate or fake pages. The ratio of internal (\textbf{f62}) and external (\textbf{63}) redirections are proposed for distinguishing phishing web pages.\\
		\noalign{\smallskip}\hline\noalign{\smallskip}
		\textbf{f64-65} & Ratio Internal/External errors & Float & IC1 & Fake hyperlinks are usually present in phishing web pages. Therefore, all hyperlinks of web pages are checked and the ratio of internal (\textbf{f64}) and external (\textbf{f65}) hyperlinks connection errors are counted.\\
		\noalign{\smallskip}\hline\noalign{\smallskip}
		\textbf{f66} & Login forms & 0/1 & IC2 &Login forms are another means commonly used for stealing information of web users. Login forms with external action links or empty actions are considered phishing. Empty action formats considered in this study are: "", "\texttt{\#}", "\texttt{\#nothing}", "\texttt{\#doesnotexist}", "\texttt{\#null}", "\texttt{\#void}", "\texttt{\#whatever}", "\texttt{\#content}", "\texttt{javascript::void(0)}", "\texttt{javascript::void(0);}", "\texttt{javascript::;}", "\texttt{javascript}".\\
		\noalign{\smallskip}\hline\noalign{\smallskip}
		\textbf{f67} & External favicon & 0/1 & IC1 & To mimic legitimate websites, phishers use the same favicon icon of the target website in the address bar of navigators. Therefore, websites using external favicons are considered phishing.\\
		\noalign{\smallskip}\hline\noalign{\smallskip}
		\textbf{f68} & Links in tags & Float & IC1 & In legitimate websites it is expected that \texttt{<Link>} tags use links pointing to web pages of the same domain as the URL. Therefore, the ratio of internal links in \texttt{<Link>} tags is considered for phishing detection.\\
		\noalign{\smallskip}\hline\noalign{\smallskip}
		\textbf{f69} & Submit to Email & 0/1 & IC2 & Phishers submit user inputs in web forms into specific email addresses. Form actions containing '\texttt{mailto:}' or '\texttt{mail()}' are considered phishing.\\
		\noalign{\smallskip}\hline\noalign{\smallskip}
		\textbf{f70-71}& Ratio internal/external media & Int & IC1 & Legitimate websites mostly use media (images, audio, video) stored in the same domain. Phishing websites use more external media, usually stored in the target website domain, to save the storage space. Ratios of internal (\textbf{f70}) and external (\textbf{f71}) media file links are counted and used for distinguishing legitimate form phishing websites. \\
		\noalign{\smallskip}\hline\noalign{\smallskip}
		\textbf{f72}& SFH & 0/1 & IC2 &Normally, actions should be taken upon submitted information on web page forms. Therefore, Forms with an empty string or '\texttt{about:blank}' are considered parts of phishing web pages.\\
		\noalign{\smallskip}\hline\noalign{\smallskip}
		\textbf{f73} & Invisible iframe & 0/1  & IC2 & Frame tags are used to incorporate additional web pages to those actually shown. Phishing websites may use \texttt{<iframe>} with invisible border so that users may think that additional pages are part of current websites while they are actually from different domains. Therefore, the use of invisible \texttt{<iframe>} tags is considered as a phishing indicator.\\
		\noalign{\smallskip}\hline\noalign{\smallskip}
		\textbf{f74}& Pop-up ~~~~~~ window & 0/1 & IC2 & Pop-up windows are used by legitimate websites to alert users with warnings but rarely used to submit user information. The presence of pop-up windows with text fields is considered as a phishing indicator.\\
		\noalign{\smallskip}\hline\noalign{\smallskip}
		\textbf{f75}& Safe anchor & Int & IC2 & The \texttt{<a>} tag is used to enable linking from one page to another. Tags with one of the  following links $\{$'\texttt{\#}', '\texttt{javascript}', '\texttt{mailto}'$\}$ are considered unsafe. Thus, we consider the number of unsafe anchors.\\
		\noalign{\smallskip}\hline\noalign{\smallskip}
		\textbf{f76-77} & Right-click & 0/1 & IC2 & Scripts can be used to disable the right-click function. However, this can also be used by phishers to unable viewing the source code of web pages. Therefore, the presence of \texttt{onmouseover}' attribute (\textbf{f76}) and use of '\texttt{event.button==2}' as an action to '\texttt{onmouseover}' attribute (\textbf{f77}) is considered as a phishing indicator.\\
		\noalign{\smallskip}\hline\noalign{\smallskip}
		\textbf{f78} & Empty title & 0/1 & IC2 & Most legitimate websites describe the title of web pages in the \texttt{<title>} tag. The absence of web page title is considered as phishing indicator.\\
		\noalign{\smallskip}\hline\noalign{\smallskip}
		\textbf{f79} & Domain in title & 0/1 & IC2 & Legitimate websites often use the domain name as part of the title of web pages. Phishing websites use legitimate domains in titles to deceive users. Therefore, the presence of the domain of URL as part of the web page title is considered as legitimacy indicator.\\
		\noalign{\smallskip}\hline\noalign{\smallskip}
		\textbf{f80}& Domain within ~~~~~~~ copyright & 0/1 & IC2 & Legitimate websites indicate their domain name within the copyright logo. Phishing websites do not use their actual domain. Presence of the domain of URLs within the copyright logo is a legitimacy indicator.\\
		\noalign{\smallskip}\hline
	\end{longtable}
\end{small}

\subsection*{External-based features} External features (\textbf{E}) are obtained by querying reference third party services and search engines. Examples of such third party services are WHOIS\footnote{WHOIS service: \url{https://www.domain.com/whois}}, Alexa, Openpagerank\footnote{Openpagerank website: \url{https://openpagerank.com}} and Google\footnote{Google search engine: \url{https://www.google.com}}. The list of external-based features experimented in this study are given in Table~\ref{tab:tptfe}.








\begin{small}
	\begin{longtable}{l|p{.7in}|l|c|p{3.9in}}
		\caption{External-based features for website phishing detection}
		\label{tab:tptfe}       
		\endfirsthead
		\endhead
		\hline\noalign{\smallskip}
		Index & Feature & Type & Class & Intuition\\
		\noalign{\smallskip}\hline\noalign{\smallskip}
		\textbf{f81} & WHOIS registered domain & 0/1 & E & Domains of phishing websites do not match any WHOIS database record contrary to most legitimate domains. Therefore, URLs with domains not registered in WHOIS are considered phishing.\\
		\noalign{\smallskip}\hline\noalign{\smallskip}
		\textbf{f82}  & Domain registration length & Int & E & Phishing websites live for a short period of time, while legitimate websites are regularly paid for several years in advance. Instead of proposing a specific threshold as proposed in~\cite{Rajab2017,Nagaraj2018,Zaini2019}, we use the number of years the domain renewal amount was paid as a phishing indicator.\\
		\noalign{\smallskip}\hline\noalign{\smallskip}
		\textbf{f83} & Domain age & Int & E & Since phishing websites are short lived, the age of URL domains is considered as a phishing indicator.\\
		\noalign{\smallskip}\hline\noalign{\smallskip}
		\textbf{f84} & Web traffic & Int & E & Phishing websites generally have less number of visitors compared with legitimate websites. Alexa is used to identify the web traffic of URLs.\\
		\noalign{\smallskip}\hline\noalign{\smallskip}
		\textbf{f85}  & DNS record & 0/1 & E & Domain Name Server (DNS) is mandatory to retrieve the IP address of URLs for access. Therefore, URL domains must be registered within the DNS. A missing DNS recored is a phishing indicator.\\
		\noalign{\smallskip}\hline\noalign{\smallskip}
		\textbf{f86} & Google index & 0/1 & E & Phishing websites live for short times and are often accessible through direct links sent to users in emails, they do not need to be indexed by Google. Web pages not indexed by Google are supposed phishing.\\
		\noalign{\smallskip}\hline\noalign{\smallskip}
		\textbf{f87} & Page rank & Int & E & Phishing web pages are not very popular, hence, they suppose to have low page ranks compared with legitimate web pages. We use Openpagerank to get the value of this feature.\\
		\noalign{\smallskip}\hline
	\end{longtable}
\end{small}

\subsection{Dataset generation}
Features are extracted for each URL making use of information stored in a dataset indexed by the URL field  (\textit{dataset A} in Figure~\ref{fig:dataset}). \textit{Dataset A} is stored  in pickle format where each row contains an URL, its source (i.e., Yandex, Alexa, Phishtank, Openpish) and its generated DOM tree object. Generated feature vectors are stored in a separate dataset indexed by URLs in CSV format (\textit{dataset B} in Figure~\ref{fig:dataset}). 
Thereafter, \textit{dataset B} is balanced with 50\% for each class (i.e., legitimate, phishing). Finally, rows are shuffled and the dataset is made ready to use. The collection process was elaborated on March 2020 and the source code for features extraction and final datasets A and B are made available to users for replication and enhancement. 

\section{Experiments}
\label{sec:process}
In this study we perform five experiments on the collected dataset. The aim of the two first experiments is to identify best classifier(s) for website phishing detection, where the aim of the two next experiments is to investigate the role of feature selection methods in improving the performance of classifiers. Finally, the last experiment aims to identify features and/or feature classes suitable for instant phishing detection. For this last experiment, we use Python scripts to estimate the average extraction time of features. 

We conduct the first four experiments within the Weka platform~\cite{Frank2010} without any modification in the core development of the tool. Therefore, default parameter values associated to each used algorithm are adopted. For the experiments, we make use of the 10-fold cross-validation method~\cite{Picard1984, Gunasegaran2017} to measure the predictive ability of classifiers. The adoption of this method aims to minimize the bias produced by random sampling of train and test data samples.
%
%
For performance evaluation, we use two main metrics:

\begin{enumerate}
	\item \textit{Accuracy}: represents the ratio of correct predicted samples to the total number of samples. Accuracy metric works well for balanced datasets which is the case of the dataset used in this study. The accuracy of a model is calculated using the following formula:
	
	\begin{eqnarray*}
		\text{Accuracy} &=& \frac{\text{TP} + \text{TN}}{\text{TP} + \text{TN}+ \text{FP} + \text{FN}}\\
	\end{eqnarray*}
	
	True Positive (TP) designates the number of correct predictions of phishing web pages where True Negative (TN) designates the number of correct predictions for legitimate web pages. Similarly, False Positive (FP) designates the number of incorrect predictions of phishing web pages where False Negative (FN) designates the number of incorrect predictions of legitimate web pages

	%
	
	\item \textit{Macro F1-score:} captures the mean of class-wise F1-scores. Macro F1-score is obtained by averaging  F1-scores computed for each class $i$. Macro F1-score is calculated as described in the following formula:
	
	\begin{eqnarray*}
		\text{Macro F1-score} & = & \frac{1}{N} \sum_{i=0}^{N} {\text{F1-score}_i}
	\end{eqnarray*} 
	
	F1-score for each class represents the best trade-off between precision and recall of the class. It is calculated using the following formula:
	
	\begin{eqnarray*}
		\text{F1-score}_i & = & 2 * \frac{\text{Precision}_i * \text{Recall}_i}{\text{Precision}_i + \text{Recall}_i}
	\end{eqnarray*} 
	
	Recall and precision are calculated in terms of TP, FP and FN as follows:
	
	\begin{eqnarray*}
		\text{Recall} &=& \frac{\text{TP}}{\text{TP} +\text{FN}}\\
		\text{Precision}&=& \frac{\text{TP}}{\text{TP} +\text{FP}}\\
	\end{eqnarray*} 	
\end{enumerate}

\subsection{Classifier evaluation}
The aim of the experiments of this part is to identify the best classifier(s) for website phishing detection. Das et al.~\cite{Das2020} elaborated a ranking based on their usage frequency in previous studies. In this study, we test and compare the performance of the five best classifiers identified in~\cite{Das2020} on the collected dataset considering: (1) the different classes of features, and (2) the effect of combining best models for each class.   

\subsection*{Experiment \RN{1}}
\begin{figure}[h]
	\centering
	\includegraphics[width=.8\textwidth]{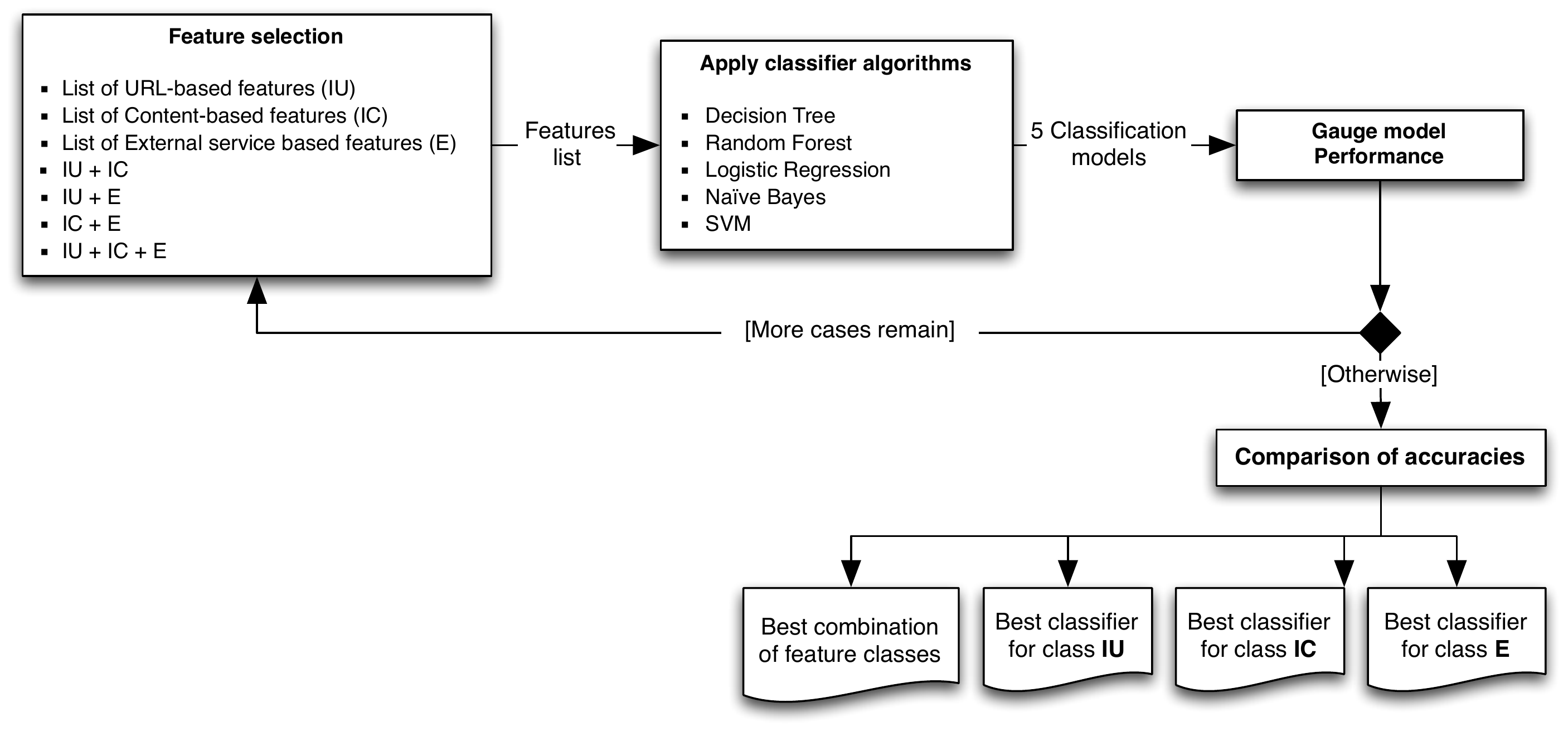}
	\caption{Identification process for best classifier(s) and best combination of feature classes.}     
	\label{fig:classifier} 
\end{figure}

In the first experiment we aim to identify the best classifier(s) for each individual class of features and best combination of feature classes. Achieving this goal, we apply several classifiers individually to different classes of features. The performance of each classifier is saved and compared with other classifiers. Best performed classifiers for each class of features are made candidates for Experiment \RN{2}. Moreover, we test all the classifiers with hybrid selection of features. Therefore, we experiment the different possible combinations of feature classes. Figure~\ref{fig:classifier} depicts the overall process adopted for this first experiment.

\subsection*{Experiment \RN{2}}
Best classifiers of each individual class identified in Experiment \RN{1} are used to check potential performance improvements that may result from their combinations. We examine four types of combinations: 

\begin{itemize}
	\item \textit{AND combination:} each one of the three classifiers makes a prediction for each test instance. A test instance is considered phishing if all the three classifiers predict it as phishing.
	\item \textit{OR combination:} each one of the three classifiers makes a prediction for each test instance. A test instance is considered phishing if at least one of the three classifiers predicts it as phishing.
	\item \textit{Stacking:} a new classifier is trained on the outputs of the three base classifiers. The final prediction is the one produced by the new classifier.
	\item \textit{Majority voting:} each one of the three classifiers makes a prediction (vote) for each test instance and the final prediction is the one that receives more than one vote. 
\end{itemize}

\begin{figure}[h]
	\centering
	\includegraphics[width=.8\textwidth]{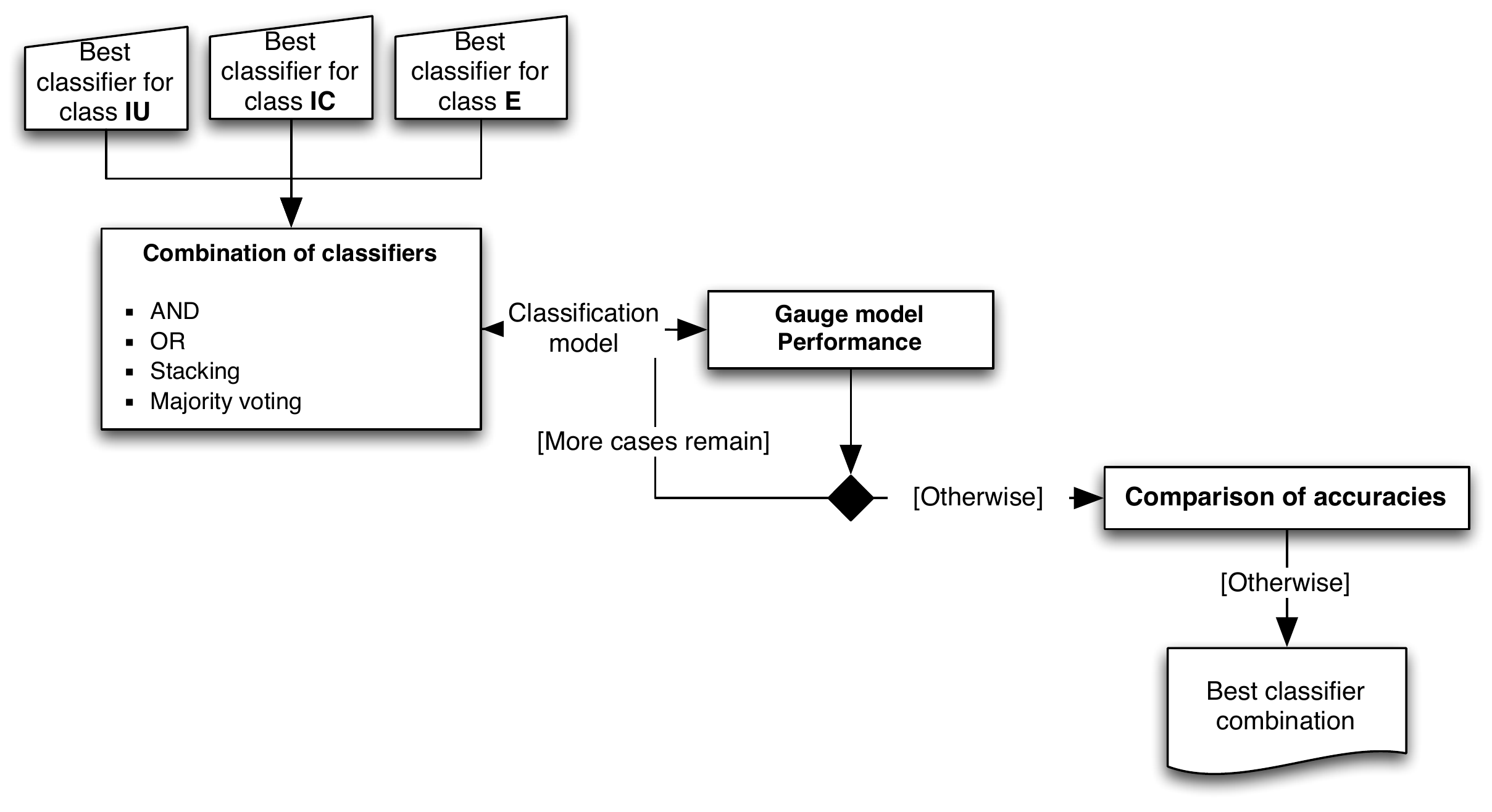}
	\caption{Combination of models trained on different classes of features.}     
	\label{fig:classifier1} 
\end{figure}

Figure~\ref{fig:classifier1} depicts the overall process adopted for this experiment.

\subsection{Feature selection methods evaluation}
The experiments of this part are concerned with the selection of best feature subsets to improve the accuracy of models. For this sake, we examine the ability of existing techniques in the selection of discriminative features, namely \textit{filter} and \textit{wrapper} methods. The aim of the experiments is to check the validity of the widely accepted hypothesis stating that higher ranked features provide higher classification accuracy. 

\subsection*{Experiment \RN{3}}
\begin{figure}[h]
	\centering
	\includegraphics[width=.8\textwidth]{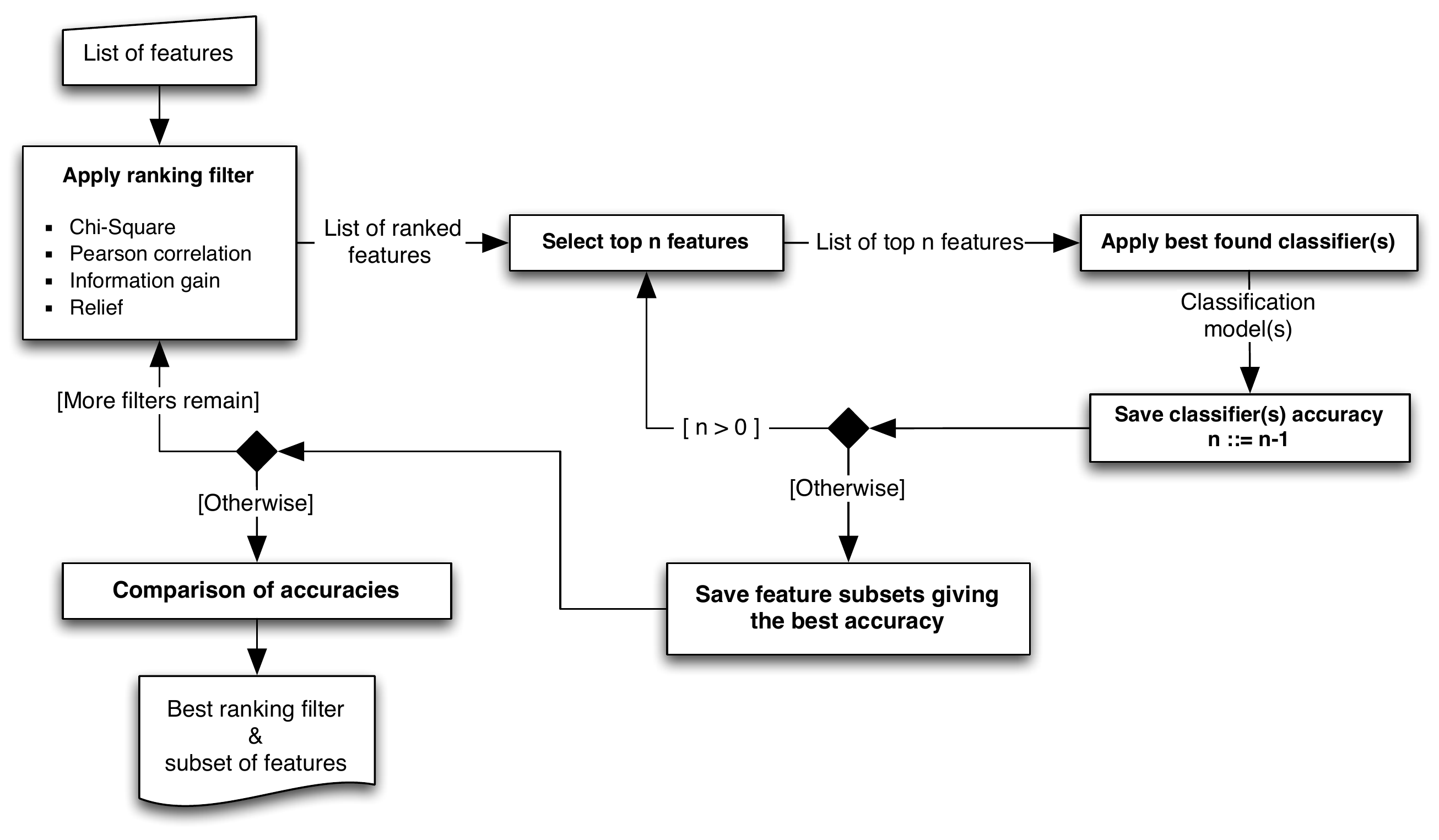}
	\caption{Identification process for best filter(s) and feature subset(s).}     
	\label{fig:slectdata} 
\end{figure}

Filter methods select features independently of any machine learning algorithm. They select features based on scores associated to features based on their correlation with the class attribute values.
In this experiment, we examine the efficiency of different filter-based algorithms in the selection of distinguishing features. For this purpose, we adopt the process depicted in Figure~\ref{fig:slectdata}. 

Four ranking filters are tested: \textit{chi-square}, \textit{Pearson correlation}, \textit{information gain} and \textit{relief}~\cite{Kira1992}.  

\begin{itemize}
	\item \textit{Chi-Square:} estimates the value of each feature by measuring the chi-squared statistic value with respect to the class attribute values.
	\item \textit{Pearson's correlation:} estimates the value of each feature by quantifying the linear dependency between the features and the class attribute values. 
	\item \textit{Information gain:} entropy-based feature evaluation method that estimates the quality of a feature by measuring the amount of information provided by the feature with respect to the class attribute value.
	\item \textit{Relief}: estimates the quality of a feature according to how well its value distinguishes between nearest instances to each other. 
\end{itemize}

Firstly, all the features are selected and a ranking filter is applied each time. Best found classifier(s) from Experiment \RN{2} are tested on ranked features. Thereafter, less important features are removed from the list one by one and classifiers are trained on the remaining features. Best subset of features is the one giving best accuracy values of applied classifiers. 

\subsection*{Experiment \RN{4}}
\begin{figure}[h]
	\centering
	\includegraphics[width=.75\textwidth]{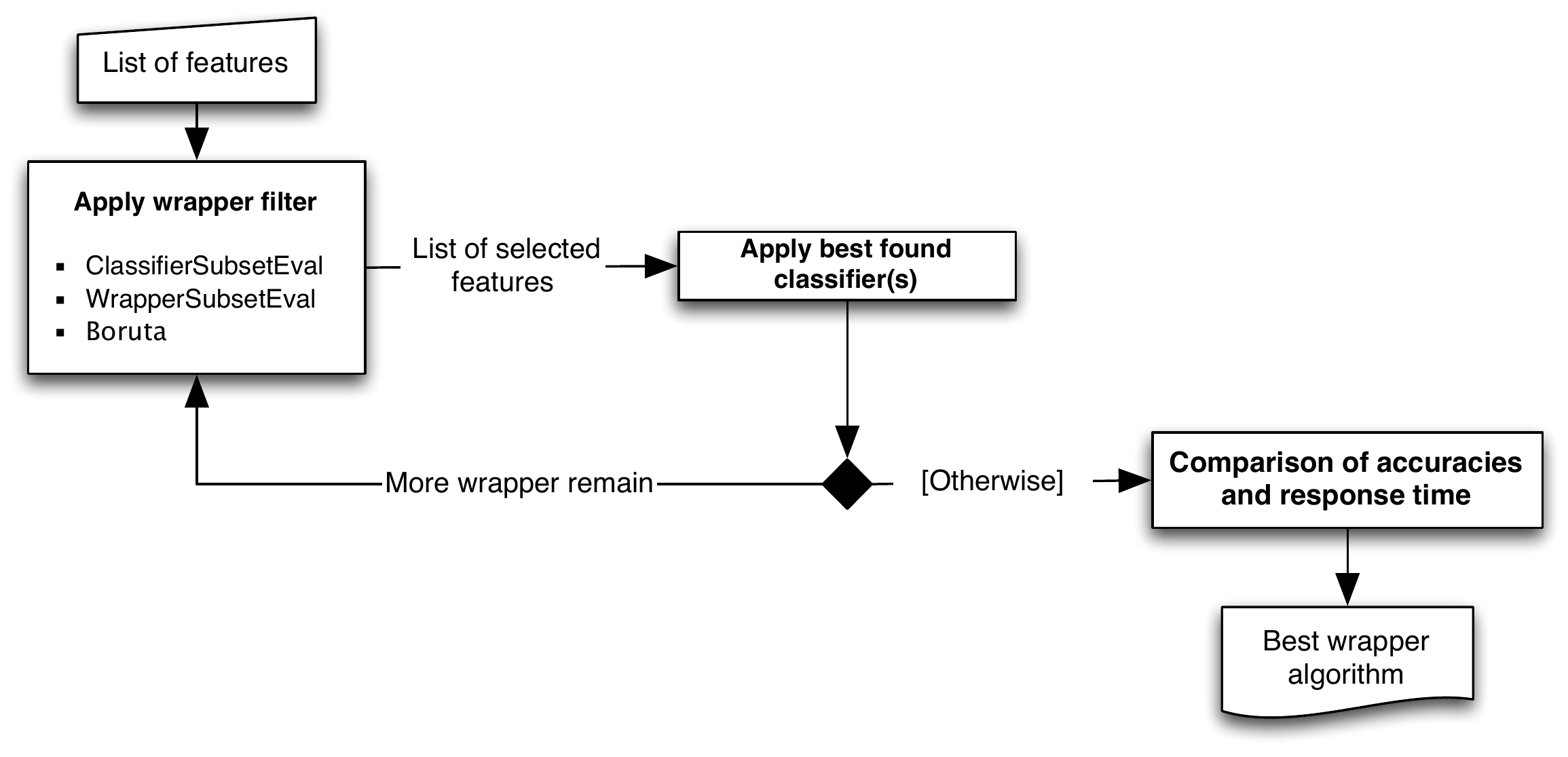}
	\caption{Identification process for the best wrapper algorithm for feature selection.}     
	\label{fig:slectdata1} 
\end{figure}

Compared with filter methods, the feature selection in wrapper methods depends on the selected machine learning algorithm. Those methods start by feeding the selected classifier with a subset of features. Based on the obtained accuracy and inferences drawn from previous computations, the algorithm decides to add or remove features from the subset until reaching the best accuracy. 
In this experiment, we focus in the evaluation of the efficiency of different wrapper-based algorithms for the selection of features. Specifically, we examine both \textit{ClassifierSubsetEval} and \textit{WraperSubsetEval} evaluators~\cite{Frank2010}. Both evaluators are applied with the \textit{set training dataset} option of Weka. Best-First search is adopted since it is found in~\cite{Kohavi1997} as the best performing method. Figure~\ref{fig:slectdata1} depicts the process adopted for this experiment. In addition, we use Python implementation of the Boruta wrapper~\footnote{Python implementation of Boruta wrapper algorithm: \url{https://pypi.org/project/Boruta/}} to check its efficiency in the selection of best features.

\subsection{Experiment \RN{5}: Feature extraction runtime analysis}
In this experiment, we examine the required time for the construction of feature vectors used as inputs for machine learning classifiers. This is needed to identify suitable features for runtime detection. For this sake, we firstly check the performance of each class individually. When notable delays are noticed, we inspect individual features in each class in the aim to identify those who may cause such delays. For the experiment, we estimate the average extraction time of all the 11430 samples of the collected dataset. Since the extraction time may differ from legitimate to phishing web pages due their different size of contents, our adoption of all the instances of the dataset is feasible due the balance nature of the dataset.

\section{Experimental results}
\label{sec:results}
In this section, we present the details of the different conducted experiments with the obtained results. In total, five experiments are performed. In the first four experiment, we evaluate the performance of used classifiers and check for potential improvements. In the last experiment, we estimate the extraction time of different features and feature classes.

\subsection{Experiment \RN{1}. Performance of classifiers trained on different classes of features}
\begin{figure}[h]
	\centering
	\includegraphics[width=.84\textwidth]{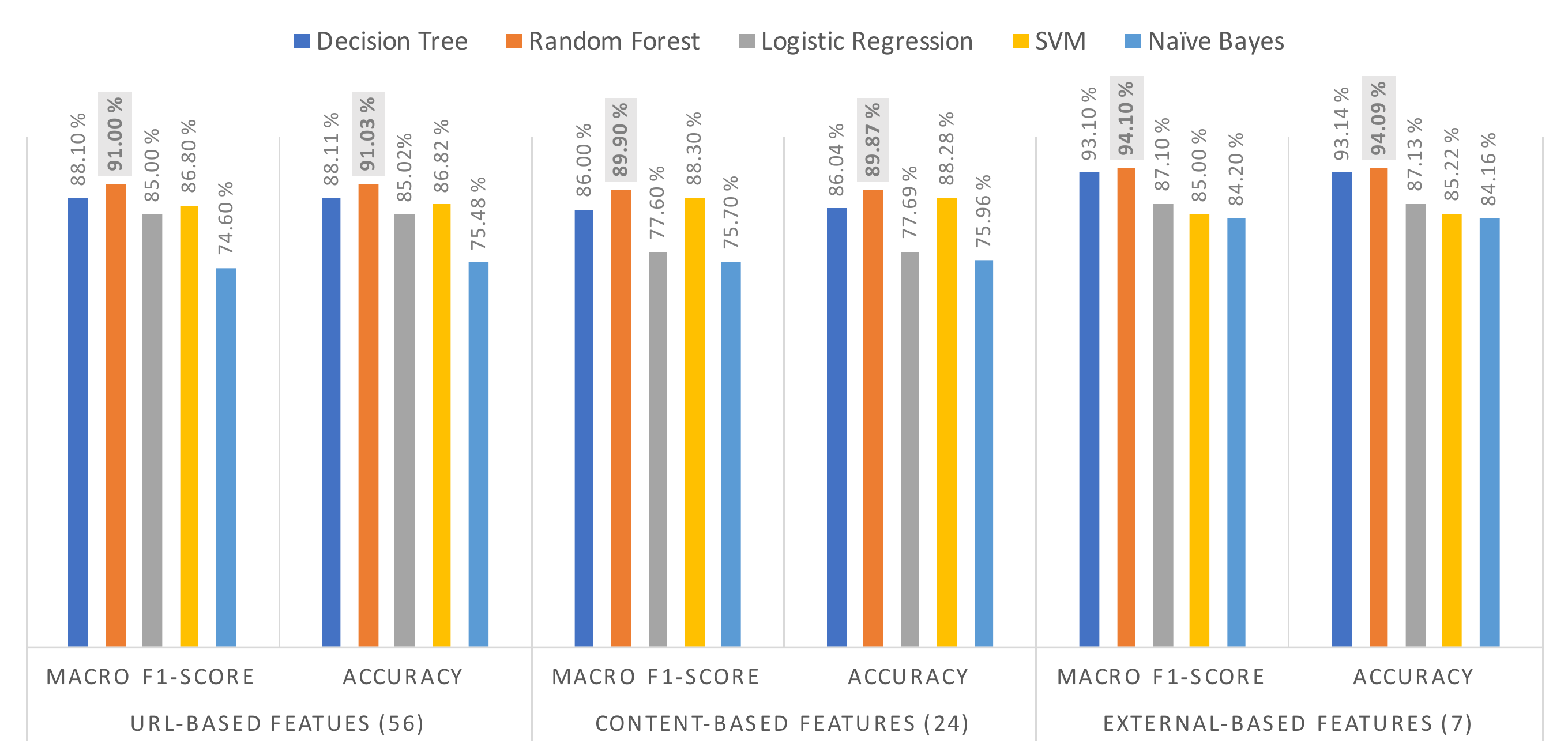}
	\caption{Performance of classifiers trained on individual class of features.}     
	\label{fig:models} 
\end{figure}

In this experiment, we evaluate the predictive ability of five common used classifiers for website phishing detection on different classes of features. Decision Tree, Random Forest, Logistic Regression, Naïve Bayes and SVM are found in~\cite{Das2020} as the most frequently used classifiers by contemporary studies. Those same classifiers are used in this experiment and trained individually on each class of features. 
Figure~\ref{fig:models} shows the experimental results of the examined classifiers on URL, content and external-based features. The results clearly show that Random Forest outperforms all the examined classifiers with higher values of both accuracy and Macro F1-score. This, in fact, is consistent with the results obtained by earlier studies that claim that Random Forest is the best classifier in term of accuracy~\cite{Chiew2019, Sahingoz2019}. Contrary to the results obtained by Jain et al.~\cite{Jain:2019}, logistic regression classifier provides the third best accuracy score for content-based features regarding the collected dataset. Jain et al.~\cite{Jain:2019} used features \textbf{f57}-\textbf{f67} that are also experimented separately and gave an accuracy score of 77.11\% within logistic regression which is less than that obtained by Random Forest classifier 86.46\%. This clearly indicates that the results of Jain et al.~\cite{Jain:2019} are dataset dependent.

The results also show that external-based features provide the best accuracy score 94.09\% within Random Forest classifier. URL-based features come in the second place with an accuracy score 91.03\%, where content-based features provide the less accuracy score 89.87\%. This also invalidate the results obtained by Choo et al. in~\cite{Choo2016} where it was found that URL-based features are more important than the other two classes of features. Choo et al. in~\cite{Choo2016} used only three external-based features \textbf{f83-85}. Using conceptual replication of the work of Choo et al.~~\cite{Choo2016} within the collected dataset, we obtained an accuracy score of 86.42\% within Random Forest classifier which is much more less that that obtained by including all external based features.

\begin{figure}[h]
	\centering
	\includegraphics[width=1.02\textwidth]{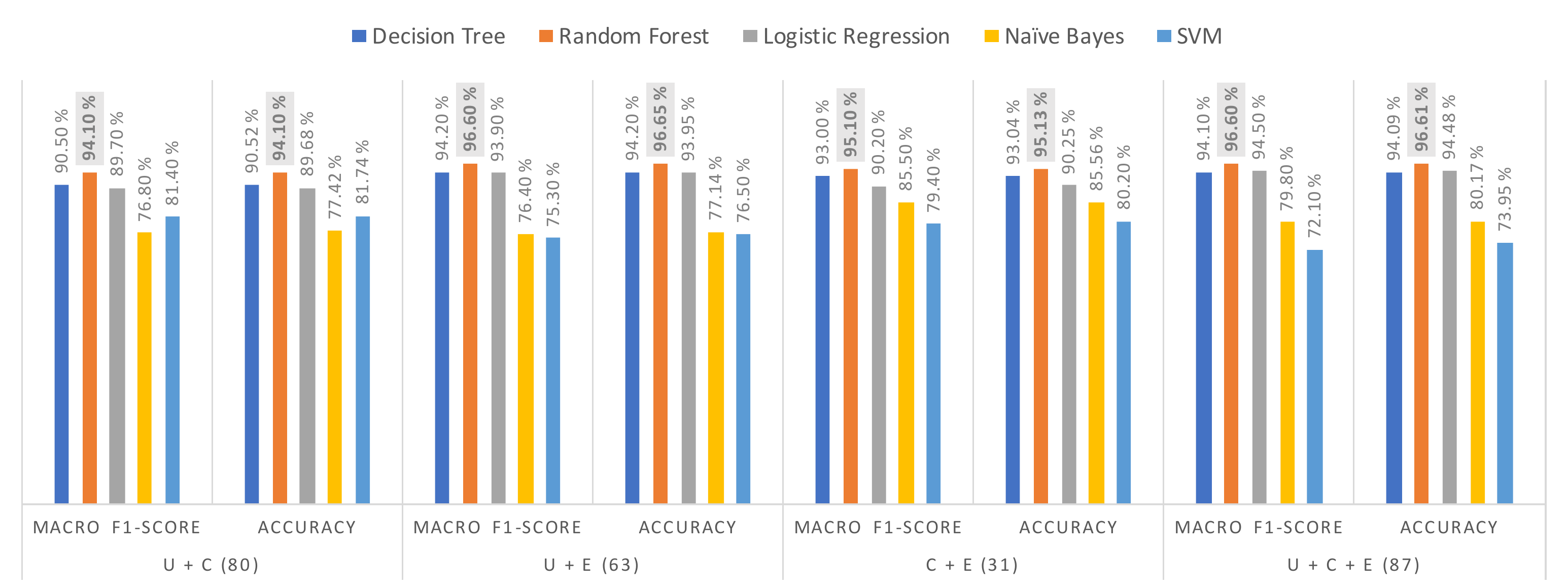}
	\caption{Performance of classifiers trained on pairwise combined class of features.}     
	\label{fig:models1} 
\end{figure}

Regarding the obtained results using individual class of features, it can be concluded that none of the classes has the potential of being used independently for website phishing detection. Therefore, we examine the effect of combining features from different classes in improving the performance of classifiers. We start by pairwise combination of classes and we end with the combination of all the features. Figure~\ref{fig:models1} shows the results of the experiment. The results confirm again that Random Forest outperforms all the other classifiers and provides higher accuracy scores. The performance of the Random Forest classifier is increased in all the cases where the combination of URL and external-based features provide the higher accuracy score 96.60\%. This was expected since these two classes provided the higher scores when they are considered individually. The combination of content with external based features also improved the performance with 1.04\% compared with considering merely external-based features. Surprisingly, the combination of all the features gave the second best accuracy sore 96.61\% but with the exact same Macro F1-score 96.60\%. 

As can be noticed, Random Forest outperforms all the other classifiers in all cases but provides the highest scores within hybrid features. Moreover, not all the classes are equally important; content-based features are found less important and external based features gave about the same score obtained when combining URL and content-based features (i.e., 96.09\% against 96.10\%). This advocates the use of Random Forest classifier with hybrid features instead of individual classes. Moreover, Decision Tree comes in the second place except when all the features are considered, where Logistic regression comes close to Random Forest. SVM classifier gave the worst performances except in the case of combining URL and content-based features.


\subsection{Experiment \RN{2}. Performance of combined models each trained on a different class of features}
\begin{figure}[h]
	\centering
	\includegraphics[width=.65\textwidth]{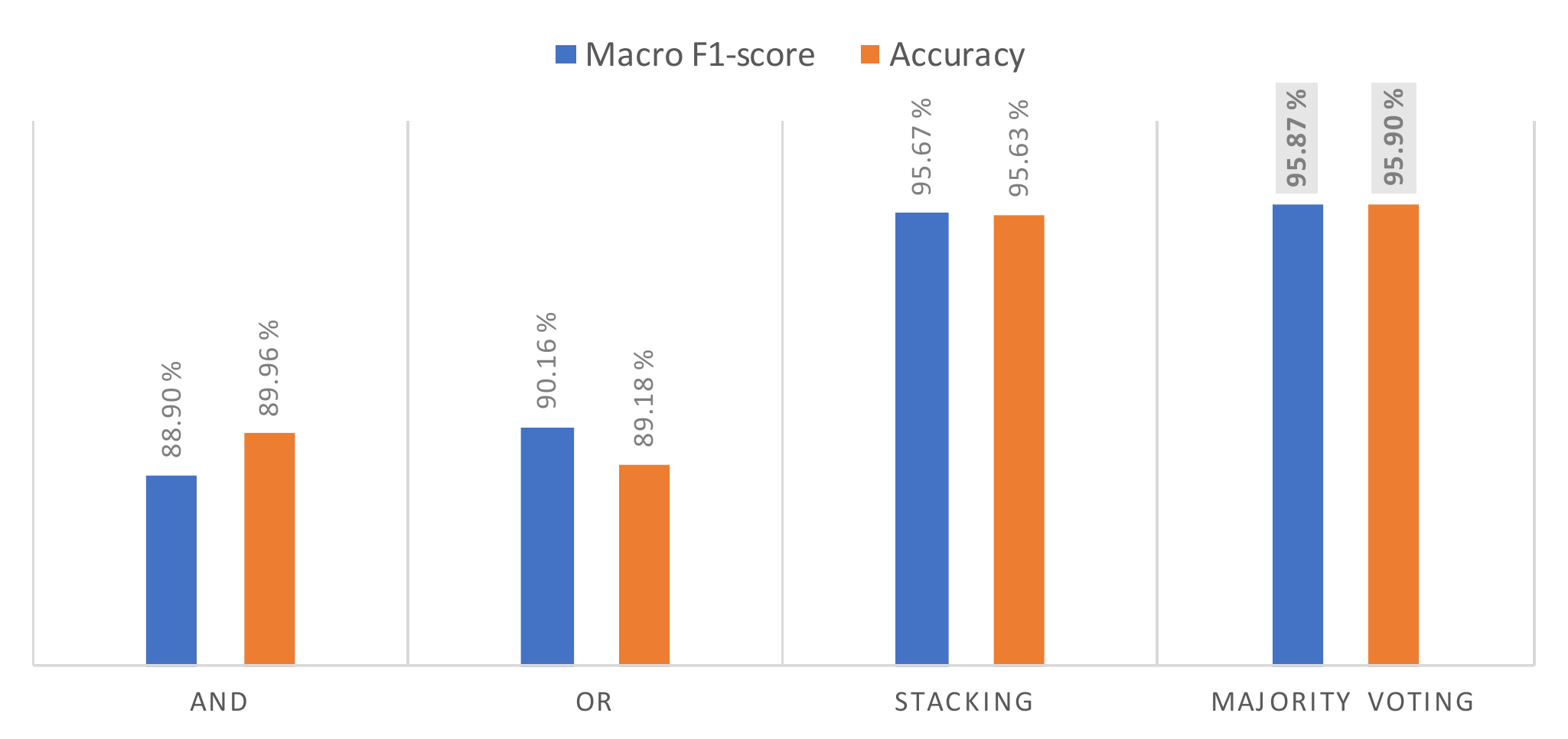}
	\caption{Performance of combined models.}     
	\label{fig:combination} 
\end{figure}

Following the process given in Figure~\ref{fig:classifier1}, we examine, in this experiment, the combination of three Random Forest models each trained on one class of features aiming to improve predictions accuracy. AND, OR, stacking and majority voting combinations are examined. In each case, the final prediction for test instances depends on the prediction of the three base models.
The results presented in Figure~\ref{fig:combination} show that none of the combinations improves the accuracy. However, stacking and majority voting gave approximate scores to those obtained by using a single model trained on content assembled with external-based features. 

\subsection{Experiment \RN{3}. Performance by training on features selected using filter ranking methods}
\label{sec:franks}
In this experiment, we evaluate the accuracy of the best found classifier (i.e., Random Forest) on decremental selection of the features set. Features, are first ranked using chi-square, Pearson correlation, information gain and relief filters. Thereafter, the Random Forest classifier performance is evaluated on selected features where the lowest ranked features are eliminated one by one in each step. 
Table~\ref{tab:ranks} shows the list of the top 25 ranked features using the different examined filters with respect to the class of web page URLs. From the table, \textit{google\_index}~(\textbf{f86}) is the top feature identified by all the filters and 7 features (i.e; \textbf{f86}, \textbf{f87}, \textbf{f21}, \textbf{f26}, \textbf{f79}, \textbf{f83}, \textbf{f58}) are found in the top 25 features of all the filters. Examining only those 7 features we reached a maximum accuracy score 94.67\% with Random Forest. 
Moreover, the top 25 features of all the filters come from the three different classes which justifies the need for using hybrid features for the identification of phishing web pages. Specifically, \textbf{IU2} is found to be the best class where the presence of their incorporated features into the top 25 lists varies from 20\% to 60\%. In the contrary \textbf{IC2} is found to be the worst class since the presence of their features in to top 25 lists varies only from 8\% to 12\%. 

\begin{table}[h]
	\centering
	\scriptsize{
		\caption{Top 25 ranked features per each used filter}
		\label{tab:ranks}       
		\begin{tabular}{rll|ll|ll|ll}
			\cline{2-9}\\
			&\multicolumn{2}{c}{Pearson correlation}& \multicolumn{2}{|c}{Information gain}&\multicolumn{2}{|c}{Relief}& \multicolumn{2}{|c}{Chi-Square} \\
			\hline\noalign{\smallskip}
			Order &Feature & Class & Feature & Class & Feature & Class& Feature & Class \\
			\noalign{\smallskip}\hline\noalign{\smallskip}
			1 & \textbf{f86} & E & 	\textbf{f86} & E & \textbf{f86} & E& \textbf{f86} & E\\
			2& \textbf{f87} & E &  \textbf{f84} & E & \textbf{f87} & E &  \textbf{f84} & E\\
			3& \textbf{f21} & IU2 & \textbf{f57} & IC1 & \textbf{f21} & IU2& \textbf{f57} & IC1\\
			4& \textbf{f26} & IU2 & 	\textbf{f87} & E & \textbf{f83} & E &  \textbf{f87} & E\\
			5 & \textbf{f79} & IC2 &  \textbf{f83} & E & \textbf{f3} & IU1& \textbf{f83} & E\\
			6 & \textbf{f57} & IC1 &  \textbf{f59} & IC1 & \textbf{f75} & IC2&  \textbf{f59} & IC1\\
			7 & \textbf{f51} & IU2 &\textbf{f58} & IC1 & \textbf{f33} & IU2 &  \textbf{f58} & IC1\\
			8 & \textbf{f83} & E &\textbf{f75} & IC2 & \textbf{f79} & IC2 & \textbf{f75} & IC2\\
			9 & \textbf{f3} & IU1 &\textbf{f82} & E& \textbf{f71} & IC1& \textbf{f21} & IU2\\
			10 & \textbf{f7} & IU2 &\textbf{f47} & IU2& \textbf{f80} & IC2 &\textbf{f63} & IC1\\			
			11 & \textbf{f1} & IU2 & 	\textbf{f21} & IU2 & \textbf{f68} & IC1& \textbf{f82} & E\\
			12& \textbf{f58} & IC1 &  \textbf{f63} & IC1 & \textbf{f25} & IU1 &  \textbf{f47} & IU2\\
			13& \textbf{f14} & IU2 & \textbf{f68} & IC1 & \textbf{f59} & IC1& \textbf{f68} & IC1\\
			14& \textbf{f2} & IU2 & 	\textbf{f26} & IU2 & \textbf{f70} & IC1 &  \textbf{f26} & IU2\\
			15 & \textbf{f10} & IU2 &  \textbf{f51} & IU2 & \textbf{f58} & IC1& \textbf{f41} & IU2\\
			16 & \textbf{f27} & IU2 &  \textbf{f41} & IU2 & \textbf{f31} & IU1&  \textbf{f43} & IU2\\
			17 & \textbf{f43} & IU2 &\textbf{f45} & IU2 & \textbf{f34} & IU1 &  \textbf{f51} & IU2\\
			18 & \textbf{f34} & IU1 &\textbf{f43} & IU2 & \textbf{f52} & IU1$^{*}$ & \textbf{f45} & IU2\\
			19 & \textbf{f47} & IU2 &\textbf{f50} & IU2& \textbf{f67} & IC1& \textbf{f50} & IU2\\
			20 & \textbf{f31} & IU1 &\textbf{f1} & IU2& \textbf{f38} & IU2 &\textbf{f65} & IC1\\
			21 & \textbf{f78} & IC2 & 	\textbf{f49} & IU2 & \textbf{f26} & IU2& \textbf{f79} & IC2\\
			22& \textbf{f4} & IU2 &  \textbf{f2} & IU1 & \textbf{f36} & IU1$^{*}$ &  \textbf{f1} & IU2\\
			23& \textbf{f45} & IU2 & \textbf{f65} & IC1 & \textbf{f84} & E& \textbf{f70} & IC1\\
			24& \textbf{f50} & IU2 & 	\textbf{f4} & IU1 & \textbf{f7} & IU2 &  \textbf{f4} & IU2\\
			25 & \textbf{f49} & IU2 &  \textbf{f79} & IC2 & \textbf{f63} & IC1& \textbf{f49} & IU2\\	
			\hline\noalign{\smallskip}
	\end{tabular}}
\end{table}

\begin{figure}[h]
	\centering
	\includegraphics[width=.9\textwidth]{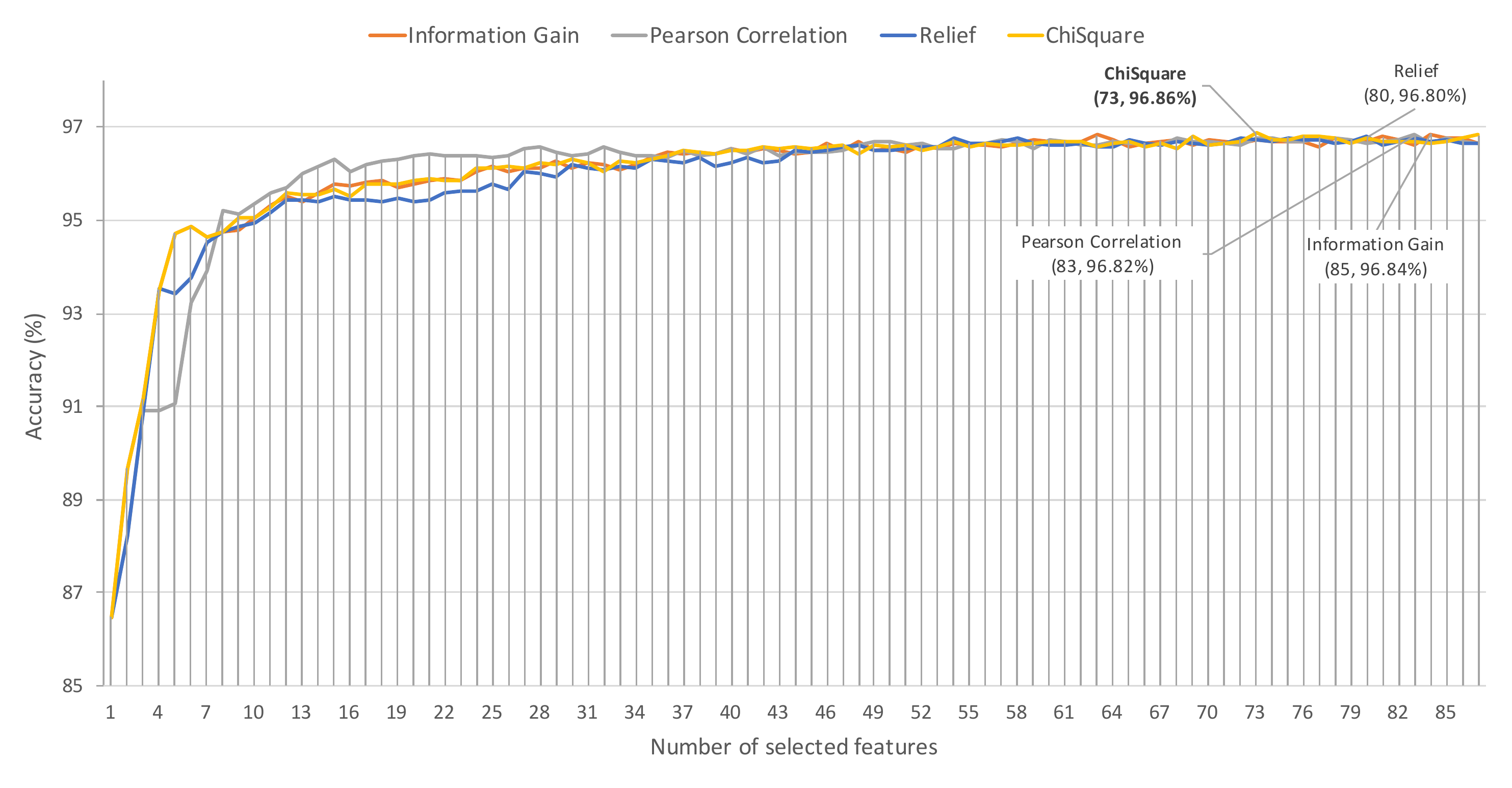}
	\caption{Accuracies of random forest prediction model with respect to stepwise feature selection.}     
	\label{fig:accuracy} 
\end{figure}

Figure~\ref{fig:accuracy} shows the accuracy graphs with respect to selected features based on the ranking results of the different filters. The features are selected such that the lowest ranked features are eliminated first and one by one in each step. The first value in Figure~\ref{fig:accuracy} designates the accuracy of the model when only the top ranked feature is selected, where the last value designates the accuracy when all the 87 features are selected. The results show that maximum accuracy rate 96.83\% is reached by selecting 73 features using chi-square filter. This value is slightly reduced by adding lower ranked features. 

From these results, it can be concluded that the use of ranking filters improves the accuracy of Random Forst. Specifically, chi-square filter gives better accuracy than other filters with less number of features. Table~\ref{tab:filters} shows the set of less important features detected by each filter. The results show that content-based features are less important since 37.50\% of these features were detected less important by three filters with a single common feature (i.e.; \textbf{f10}). Moreover, all external service based features are found important which indicates the efficiency of such class of features in distinguishing phishing web pages.


\begin{table}[h]
	\centering
	\scriptsize{
		\caption{List of less important features identified by used ranking filters}
		\label{tab:filters}       
		\begin{tabular}{llll}
			\hline\noalign{\smallskip}
			Filter & IU & IC & E\\
			\noalign{\smallskip}\hline\noalign{\smallskip}
			Pearson Correlation& \textbf{f10} & \textbf{f60}, \textbf{f64}, \textbf{f69} & None\\
			Information gain& \textbf{f10}, \textbf{f29} & \textbf{f60} & None\\
			Relief& \textbf{f19}, \textbf{f18}, \textbf{f28}, \textbf{f10} & \textbf{f69}, \textbf{f72}, \textbf{f62}  & None\\
			ChiSquare& \textbf{f10}, \textbf{f20}, \textbf{f35}, \textbf{f29}, \textbf{f37} & \textbf{f69}, \textbf{f62}, \textbf{f72}, \textbf{f64}, \textbf{f77}, \textbf{f60},  \textbf{f73}, \textbf{f66}, \textbf{f76}& None\\
			\hline\noalign{\smallskip}
			\textbf{Total (De-duplicated)} & 8 & 9 & 0\\
			\textbf{Ratio} & 14.29\% & 37.50\% & 0\%\\
			\noalign{\smallskip}\hline
	\end{tabular}}
\end{table}

\subsection{Experiment \RN{4}. Performance by training on features selected using wrapper methods}
In this last experiment, we examine wrapper methods effect on the selection of features. Wrapper methods are time expensive since they involve training a selected model using different subsets of features. However, such methods are recognized to be more efficient in the selection of discriminative features and their ability to make models more prone to overfitting. For such reasons, we decided to evaluate the efficiency of wrapper methods in improving the accuracy of Random Forest classier regarding the collected dataset. Two wrapper evaluators provided by the Weka platform are tested: \textit{ClassifierSubsetEval} and \textit{WrapperSubsetEval}. The former evaluator assesses the worth of feature subsets by their level of consistency with the class values when the training instances are projected onto the subset of features. The latter, uses the cross validation technique to estimate the accuracy of the classifier for each subset of features~\cite{Witten2011}. We also examine the Boruta wrapper algorithm~\cite{Kursa2010}. This latter finds the importance of features by creating shadow features. Specifically, the Boruta algorithm focuses on finding all the features having impacts on the classifier prediction, rather than finding subsets causing minimal errors on the performance of the classifier. Table~\ref{tab:wrapper} compares the obtained results from the three wrapper evaluators indicating the list of selected features and their performances.

\begin{table}[h]
	\centering
	\scriptsize{
		\caption{Results of the wrapper evaluators}
		\label{tab:wrapper}       
		\begin{tabular}{lcp{3in}cc}
			\hline\noalign{\smallskip}
			Evaluator & \#Selected & Features & Macro F1-score & Accuracy\\
			\noalign{\smallskip}\hline\noalign{\smallskip}
			ClassifierSubsetEval& 11 & \textbf{f49}, \textbf{f68}, \textbf{f71}, \textbf{f75}, \textbf{f81}, \textbf{f82}, \textbf{f83}, \textbf{f84}, \textbf{f85}, \textbf{f86}, \textbf{f87}& 94.80\%& 94.84\%\\    
			WrapperSubsetEval& 81 & All except \textbf{f25}, \textbf{f43}, \textbf{f65}, \textbf{f67}, \textbf{f71}, \textbf{f76}& 96.70\%&96.69\%\\  
			Boruta& 61 & All except \textbf{f9}, \textbf{f12}, \textbf{f15-17}, \textbf{f19},
			\textbf{f23}, \textbf{f28-30}, \textbf{f35}, \textbf{f37}, \textbf{f39}, \textbf{f53-54}, \textbf{f60}, \textbf{f62}, \textbf{f64}, \textbf{f66}, \textbf{f69}, \textbf{f72-74}, \textbf{f76-77}, \textbf{f81}& 96.60\%&96.64\%\\    
			\noalign{\smallskip}\hline
	\end{tabular}}
\end{table}

The results show that \textit{ClassifierSubsetEval} failed on the identification of features that improve the performance of the Random Forest classifier. \textit{WrapperSubsetEval} and \textit{Boruta} provide better and close results but Boruta is much faster than \textit{WrapperSubsetEval}. While the former took 122 seconds, the latter took about 24 hours for producing the results. However, the three evaluators fail on improving the Random Forest accuracy compared with the decremental selection of features based on filter ranking used in Experiment~\RN{3}.

\subsection{Experiment \RN{5}. Runtime analysis of feature extraction}
\begin{figure}[h]
	\centering
	\includegraphics[width=1\textwidth]{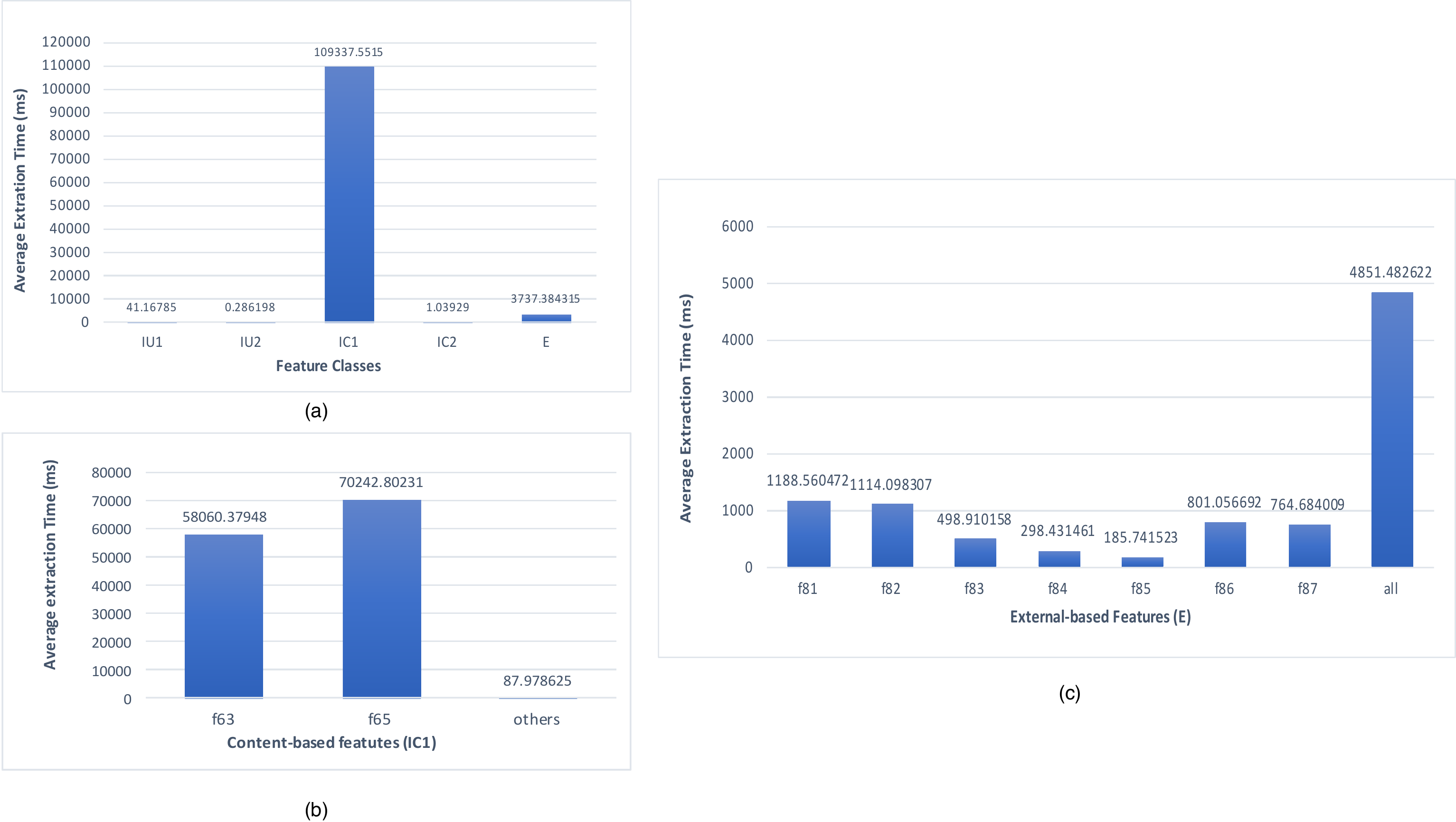}
	\caption{Average time for feature extraction}     
	\label{fig:response} 
\end{figure}

In this experiment, we examine the required time for the extraction of the different features and feature classes.
The experiment is performed in a MacBook Pro with OS X, 2.9 Ghz Intel Core i7 processor, and 8 GB of memory. Figure~\ref{fig:response} (a) shows the average time spent for the extraction of the different feature classes. 

The results of Figure~\ref{fig:response} (a) clearly indicate that both subclasses of URL-based features (\textbf{IU1}, \textbf{IU2}) and abnormal content-based features (\textbf{IC2}) are the less time consuming. Particularly, all URL-based features require about 41.5 ms for their extraction which makes them suitable enough for runtime detection systems. 

Hyperlink-based features (\textbf{IC1}) are found much slower than external-based services (\textbf{E}). For such reason we decided to go deeper in the investigation and inspect the performance of individual features of this class. Figure~\ref{fig:response} (b) shows that \textbf{f63} and \textbf{f65} are behind such delay. These two features require to check every link in the web page to find out if it is a fake or authentic; the problem with \textbf{f63} and \textbf{f65} becomes serious when web pages use a considerable number of hyperlinks and hence a great value of \textbf{f57}.  

Figure~\ref{fig:response} (c) shows that the extraction of all the external-based features (\textbf{E}) requires about 5 seconds. This can be reduced to about 4 seconds since \textbf{f81} and \textbf{f82}  come from the same service (i.e., WHOIS service) and hence they can be extracted at once through a single WHOIS service call (see Figure~\ref{fig:response} (a)).

\section{Discussion and lessons learned}
\label{sec:discussion}
Benchmark datasets for web page phishing detection are not available and their construction is crucial for fair evaluation of systems. To fill this gap, we proposed a generic process consisting of 6 main guidelines for the generation of benchmark datasets overtaking the short-lived nature of phishing websites. The proposed guidelines are deduced from best practices for ensuring homogeneity, consistency, and diversity of dataset samples. We collected a list of URLs following the proposed guidelines and we developed our own Python scripts for the extraction of 87 examined features and collected them into a unique dataset. In addition, a separate dataset in pickle format is built containing the list of DOM tree objects generated for each URL web page. This enables (1) full replication, (2) testing the dataset with more personalized features, and (3) fair comparison of systems.    
Collected datasets and Python scripts are both made publicly available at~\cite{Hannousse2020}. 

By examining the literature, we noticed that most proposed website antiphishing systems use either a single class or a limited number of features from different classes and most made choices are not well-justified. Since the aim of the proposed study is to examine the effect of a maximum number of features on an arbitrary collected dataset, 87 features from different classes are used. Some features are found not discriminative enough for the dataset. However, those features may have positive effects on other datasets. In fact, getting high-quality features is very crucial for the effectiveness of antiphishing systems and strongly dependable on used datasets. In addition, phishers are continually evolving their attack tactics bypassing existing antiphishing techniques. Therefore, higher distinguishing features in the present days may be less distinguishing or useless in the coming days. As an example, submitting stolen user information to phisher emails (\textbf{f69}) is identified as one of the strongest phishing indicators in~\cite{Rajab2017,Chiew2019}, in the collected dataset none of the 5715 phishing web pages used such technique. Another example is that the use of IP addresses (\textbf{f3}) in URL web pages is currently less used by phishers; in the collected dataset only 97 cases are identified from a set of 5715 phishing URLs. Moreover, it is stated in the recent phishing activity trends report~\cite{APWG2020} that 74\% of all phishing websites start using HTTPS protocol which may reduce the effect of \textbf{f25} in future extracted datasets. 
A hybrid strategy where features from different classes are used may alleviate this issue together with the proposed guidelines. Specifically, by testing new generated datasets following the proposed guidelines, we can keep track of the significance of features over time and upgrade systems by incorporating more discriminative features. 
Table~\ref{tab:filters} shows that all external-service based features are important where 9 from the 24 content-based features where identified to be less important. This clearly indicates the less importance of most examined content-based features regarding other classes of features. Therefore, we advocate more research on identifying important content-based features for web page phishing detection. 

Real-time detection of phishing web pages is essential for instant prevention of phishing attacks. External-based features are claimed to be the slower features in~\cite{Jain:2019} compared with content-based features. The results obtained in this study invalidate this claim. We found that hyperlink-based features (\textbf{IC1}), specifically, \textbf{f63} and \textbf{f65} are much slower than all the external-based features. Moreover, none of these features are identified as less important by examined feature selection methods except \textbf{f65} that is not selected by \textit{WrapperSubsetEval} (see Table~\ref{tab:filters} and Table~\ref{tab:wrapper}). Therefore, besides external-based services, \textbf{f63} and \textbf{f65} features are also not suitable for real-time detection. 

A combination of models trained on different classes of features are examined aiming to reduce the extraction cost. Particularly, by examining OR combination we targeted reducing the extraction cost by calling faster models. Therefore, input feature vectors of new instances required for each model are only built when their correspondent models need to be invoked. Unfortunately, the results are found discouraging where only 89.18\% of accuracy was reached. However, we obtained an acceptable accuracy rate 96.65\% by combining only URL and external service-based features (see Figure~\ref{fig:models1}). Since URL-based features can be extracted within a range of milliseconds, 4 seconds will be sufficient for building the feature vector required for such model. However an extra time is needed for the model to identify the class of the checked instance.

Moreover, as can be seen from obtained results, external features have considerable impacts on the detection of phishing web pages. We reached 94.09\% of accuracy using only 7 external service based features. For reliability purposes, only information provided by independent services such as Google and WHOIS are used in this study. Phishtank presence of URLs can also be considered as a good indicator of phishing. Phishtank service is not used in this study since it is used as a source of phishing URLs of our dataset. Adopting Phishtank presence feature may improve the performance but also causes an extra network delay.


Random Forest classifier is already identified by previous studies~\cite{Chiew2019} as the best model for web page phishing detection systems and hybrid features. The experiments conducted in this study validate this conclusion and generalize it to all classes of features. However, the order in which features are presented in the dataset has also an impact for the performance of some classifiers; particularly, Random Forest due to the incorporated mechanisms used for the selection of root and decision node features for generated trees. It is found from Experiment~\RN{3} that the use of filters for ranking features affects the accuracy of the Random Forest model. Table~\ref{tab:comparison} compares the performance of the different classifiers using feature extraction order and features ordered following the different filters. The results show that Decision tree,  Random Forest and SVM classifiers are quite sensitive to the order of features used as inputs. Moreover,  chi-square ranking improves the model accuracy up to 0.22\%. In Experiment~\RN{3}, the incremental remove of less important features improved the model accuracy up to 0.25\% with chi-square filter. 
Therefore, ranking features improve the accuracy of Random Forest models and together with decremental selection, one can get better performances than those obtained with wrapper evaluators and with less time consuming except for Boruta algorithm. 

\begin{table}[h]
	\centering
	\scriptsize{
		\caption{Sensitivity of classifiers against order of features.}
		\label{tab:comparison}       
		\begin{tabular}{lccccc}
			\hline\noalign{\smallskip}
			&Extraction order & Chi-Square order & Pearson Correlation order & Information gain order & Relief order\\
			\noalign{\smallskip}\hline\noalign{\smallskip}
			Decision Tree&94.09\%&	\textbf{94.13\%}	&94.09\%&\textbf{94.12\%}&\textbf{94.16\%}\\
			Random Forest&96.61\% & \textbf{96.83\%} & \textbf{96.65\%} & \textbf{96.76\%} & \textbf{96.66\%}\\
			Logistic regression &94.48\% & 94.48\% & 94.48\% & 94.48\% & 94.48\%\\
			Naïve Bayes &79.80\% & 79.80\% & 79.80\% & 79.80\% & 79.80\%\\
			SVM &73.95\% & 73.95\% & \textbf{73.68\%} & \textbf{73.91\%} & 73.95\%\\
			\noalign{\smallskip}\hline
	\end{tabular}}
\end{table}

\begin{figure}[h]
	\centering
	\includegraphics[width=1\textwidth]{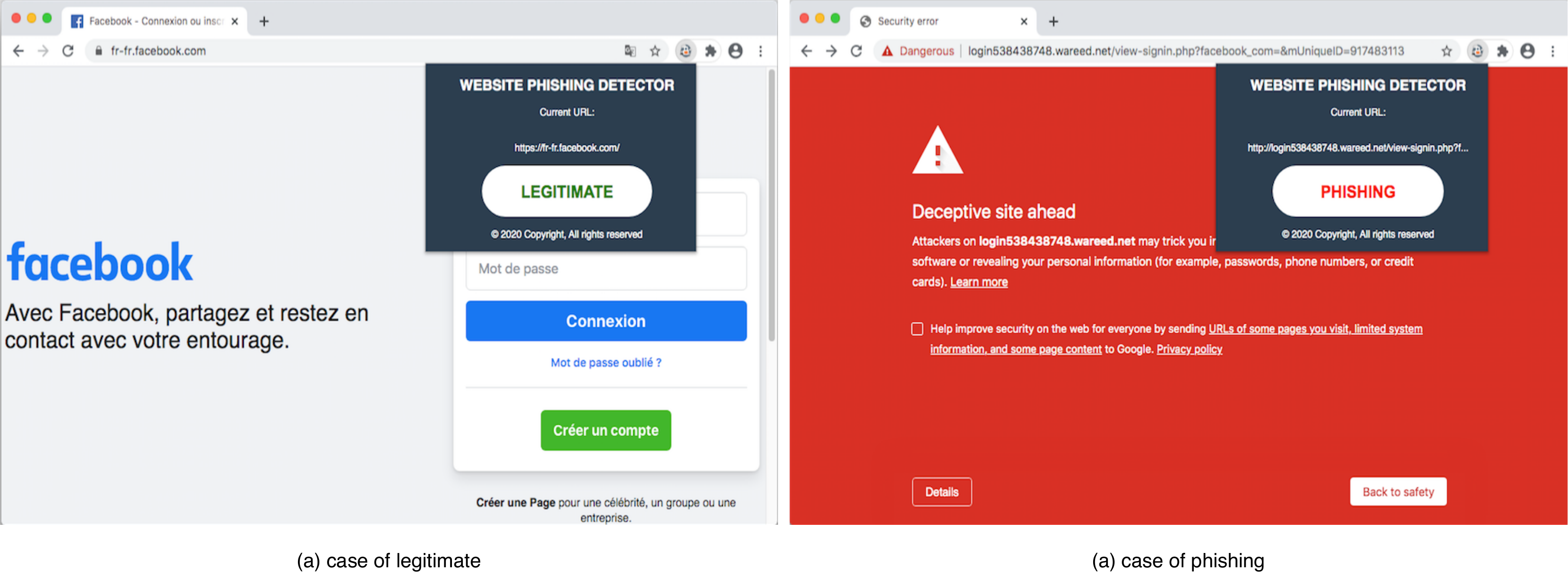}
	\caption{Screenshots of the passive plugin developed for the Chrome browser.}     
	\label{fig:plugin} 
\end{figure}

Making our results practical, we developed a passive plugin for the Google Chrome browser. The plugin opens a popup window to inform the users of the legitimacy status of the current page in the browser as depicted in Figure~\ref{fig:plugin}.  The plugin makes use of the best model found by the present study (i.e., Random forest trained on 73 features ranked following the chi-square filter). The plugin can also be used to check its efficiency regarding the active warner embedded in Google Chrome as shown in Figure~\ref{fig:plugin} (b).

\section{Conclusion}
\label{sec:conclusion}

In this study, we proposed a set of guidelines for building reproducible and extensible datasets for website phishing detection. Constructed datasets following the proposed guidelines may serve as benchmarks for machine learning based systems. A sample dataset is collected in light of the proposed guidelines and used for the examination of exiting findings. Experiments show that Random Forest classifier based systems can be used in browsers to effectively predict phishing web pages. They have comparatively higher accuracy than all the other classifiers for different feature classes. 
External service-based features despite their small number are found more effective in distinguishing phishing web pages. However, those features may cause a network delay. Examined content-based features are found the less discriminative and some hyperlink-based features may also cause severe network delays. Therefore, we advocate researchers and practitioners to scrutinize the content of phishing websites for identifying more effective content-based features. 
Since Random Forest is feature order sensitive, filter methods can effectively be used to improve the performance of the model and reduce less important features. Filter ranking together with an incremental removal of less important features provided better accuracy than wrapper methods.
Regarding the examined features, the combination of models trained on different classes of features is not capable to replace a single model trained on hybrid features. Therefore, using hybrid features provides better accuracy than using single class of features.
As a future work, we plan to validate the drawn conclusions by experimenting more datasets built in the same way as described in section~\ref{sec:data} and made as benchmarks for phishing detection. Moreover, we plan to incorporate deep learning approaches used for phishing detection in a performance analysis.

\end{document}